NOTE

# Computation capacities of a broad class of signaling networks are higher than their communication capacities

# Computation capacities of a broad class of signaling networks are higher than their communication capacities

Iman Habibi[1], Effat S. Emamian[2], Osvaldo Simeone[3] and Ali Abdi[4,5] *

[1] Iman Habibi
Center for Wireless Information Processing, Department of Electrical and Computer Engineering
New Jersey Institute of Technology, 323 King Blvd, Newark, NJ 07102, USA
Email: ih26@njit.edu

[2] Effat S. Emamian, MD
Advanced Technologies for Novel Therapeutics
Enterprise Development Center
New Jersey Institute of Technology, 211 Warren St., Newark, NJ 07103, USA
Email: emame@atnt-usa.com

[3] Osvaldo Simeone, PhD
King's Centre for Learning & Information Processing (kclip)
Center for Telecommunications Research
Department of Engineering
King's College London, London, WC2R 2LS, England, UK
Email: osvaldo.simeone@kcl.ac.uk

Ali Abdi, PhD
[4] Center for Wireless Information Processing, Department of Electrical and Computer Engineering
[5] Department of Biological Sciences
New Jersey Institute of Technology, 323 King Blvd, Newark, NJ 07102, USA
Email: ali.abdi@njit.edu

* Corresponding Author: Ali Abdi (ali.abdi@njit.edu)

**Abstract:** Due to structural and functional abnormalities or genetic variations and mutations, there may be dysfunctional molecules within an intracellular signaling network that do not allow the network to correctly regulate its output molecules, such as transcription factors. This disruption in signaling interrupts normal cellular functions and may eventually develop some pathological conditions. In this paper, computation capacity of signaling networks is introduced as a fundamental limit on signaling capability and performance of such networks. In simple terms, the computation capacity measures the maximum number of computable inputs, that is, the maximum number of input values for which the correct functional output values can be recovered from the erroneous network outputs, when the network contains some dysfunctional molecules. This contrasts with the conventional communication capacity that measures instead the maximum number of input values that can be correctly distinguished based on the erroneous network outputs.

The computation capacity is higher than the communication capacity whenever the network response function is not a one-to-one function of the input signals, and, unlike the communication capacity, it takes into account the input-output functional relationships of the network. By explicitly incorporating the effect of signaling errors that result in the network dysfunction, the computation capacity provides more information about the network and its malfunction. Two examples of signaling networks are considered in the paper, one regulating caspase3 and another regulating NFκB, for which computation and communication capacities are investigated. Higher computation capacities are observed for both networks. One biological implication of this finding is that signaling networks may have more "capacity" than that specified by the conventional communication capacity metric. The effect of feedback is studied as well. In summary, this paper reports findings on a new fundamental feature of the signaling capability of cell signaling networks.

**Introduction:** Intracellular signaling networks in a cell respond to incoming signals to regulate some target molecules, to properly control the cell function. In general, signaling networks have multiple inputs and multiple outputs. The inputs can be ligands that, upon binding to their receptors on the cell membrane, create a chain of interactions through some intermediate signaling molecules, such as receptors, kinases, phosphatases, etc. This way the network outputs, typically target proteins such as transcription factors, are collectively regulated to produce an appropriate response.

One possible way to model a signaling network is to consider it as a communication channel [1,2]. From this point of view, a signaling network communicates and conveys signals from its inputs to the outputs.

An alternative approach for modeling a signaling network is to envision it as a computing machine. In this approach, a signaling network makes some computations on the incoming signals and produces some responses at the outputs, accordingly.

Each modeling approach can reveal certain features of signaling networks. While the communication channel framework has been applied to signaling networks [1,2], the computing machine approach does not seem to have been explored so far. As demonstrated later in this paper, the developed computing machine framework appears to be advantageous for studying signaling failures and malfunctions in pathological signaling networks.

More specifically, in the introduced molecular computing machine framework, a signaling network is a molecular system that under normal conditions correctly computes the outputs based on the applied inputs. In other words, under normal conditions, a signaling network maps the inputs to outputs via a mapping or transformation $f$. Examples of this mapping for some experimentally-verified signaling networks are provided later in the paper. However, for an abnormal signaling network that contains some dysfunctional molecules due to mutations or some structural/functional abnormalities, its mapping is generally different from $f$. We call the abnormal mapping $F$. The developed computing machine approach focuses on both $f$ and $F$, and hence on comparing the ways a signaling network computes its outputs under normal and abnormal conditions.

The difference between the normal and abnormal network mappings *f* and *F*, respectively, is caused by dysfunctional molecules. This means abnormal deviation of the signaling network from its normal function, when some molecules become dysfunctional due to mutations or some structural/functional abnormalities.

A key concept in the computing machine framework is the *computation capacity*, which is fundamentally different from the communication capacity previously studied in signaling networks [1,2]. The computation capacity provides a measure of the accuracy of the computation of a desired function *f*. As demonstrated later in the paper, the computation capacity is generally larger than the communication capacity, and it directly accounts for the functional task carried out by the network, rather than focusing on input-output information transfer.

**Descriptive Comparison of Computation and Communication Capacities:** The communication capacity is the maximum amount of information that can be reliably transferred from the input of a communication channel to its output. The goal is reliable communication, i.e., correct recovery of the input message from the erroneous output. This is a reasonable model for a network that just transfers the information from its inputs to its outputs, without any processing or computation on the information. This is typically the case in man-made communication channels [3]. In such systems, the output and input are ideally the same, if there is no transmission error [3]. In contrast, in signaling networks, outputs are typically computed from inputs, and there is a desired function that maps the inputs to the outputs (examples are provided later in the paper). The goal of signaling systems therefore can be considered to be *reliable computation*, i.e., mapping inputs to correct outputs. In other words, the goal of a reliable computing network is ensuring that its erroneous outputs[1] are as close as possible to the correct outputs. This necessitates a new definition for the capacity, and the computation capacity, introduced in [4], provides a useful choice. As defined rigorously later in the paper, the *computation capacity of a signaling network is the maximum number of input values for which the functionally correct outputs, corresponding to the case when there is no dysfunctional molecule, can be recovered from the erroneous network outputs, which are affected by errors due to dysfunctional molecules*. Later in the paper we provide examples of signaling networks and calculate both their communication and computation capacities. We will show that the predictions and interpretations obtained from the computation capacity provide novel insights on signaling networks. Communication capacity is more suitable to analyze the transduction noise in a network [1,5], whereas computation capacity is well suited to study signaling failures in abnormal signaling networks.

---

[1] In a signaling network, in general input and output values are different from each other. This is because output values are computed from input values, and that is why they are different, even when there is no dysfunctional node in the network. If there is no dysfunctional node, we consider output values as correct outputs. However, if there is a dysfunctional node in the network, output values can become different from correct output values. That is why we call them incorrect or erroneous output values, or in short, erroneous outputs (resulted from errors introduced by the dysfunctional node and propagated to the network output).

The computation capacity is first introduced in the present paper for normal and abnormal networks, i.e., non-diseased and diseased (with dysfunctional molecules) networks. On the other hand, the communication capacity of non-diseased and diseased networks was previously introduced and studied in [2].

**Computation Capacity of Signaling Networks - Basic Definitions:** Consider a system such as a signaling network, with $X$ as its input, which computes the output according to the error-free mapping $f$. So, the error-free output is $f(X)$. When the system is erroneous due to the presence of some dysfunctional molecules, the mapping is called $F$, so, the erroneous output is $F(X)$. If we consider the system as a communication channel, following the regular definition of channel capacity [6], the communication capacity of this system can be written as [7]

$$C(F) = \max_{P(X)} \{H(X) - H(X|F(X))\} = \max_{P(X)} \{I(X;F(X))\}, \tag{1}$$

where the maximization is over all input distributions $P(X)$, the $H$ symbols represent entropy and conditional entropy, respectively, and $I(X;F(X)) = H(X) - H(X|F(X))$ is the mutual information between $X$ and $F(X)$. The entropy $H(X)$ measures the variability in the input, whereas the conditional entropy $H(X|F(X))$ measures the equivocation, or uncertainty, about the input given the erroneous output [6]. Following standard nomenclature, we will also refer to $R(F) = H(X) - H(X|F(X))$, for a fixed input distribution $P(X)$, as the *communication rate*. Overall, equation (1) is the commonly-used maximum mutual information between the input and output, which is well justified in the context of communication, i.e., in situations where, in the absence of electronic noise, the output is intended to reproduce the input [3]. However, in systems where the output $f(X)$ is different from the input $X$ even in the absence of any type of noise, the definition in equation (1) may underestimate the capacity of the system to reproduce the function $f(X)$. Intuitively, in computing systems where the error-free function $f(X)$ is not a one-to-one mapping, this $H(X|F(X))$ measure of equivocation generally overestimates the relevant uncertainty at the output, since the goal is recovering $f(X)$, and not the input $X$. As a result, the communication capacity may underestimate the true functional "capacity" of the system to reproduce $f(X)$. We demonstrate this later in the paper, using some experimentally-verified signaling networks. This motivates the introduction of the computation capacity concept. Note that with $Y = F(X)$ being the erroneous output, the conditional entropy or equivocation in equation (1) is given by $H(X|F(X)) = H(X|Y) = -\sum_y P(Y=y) \sum_x P(X=x|Y=y) \log_2 P(X=x|Y=y)$, where $P$ and $\log_2$ stand for probability and the base 2 logarithm, respectively. This equivocation definition is modified in what follows, to obtain a new capacity definition which is more suitable for signaling networks.

For signaling networks, a more appropriate capacity metric is one that depends on the normal network mapping function $f$ as well. We propose to use the following equivalent definitions for signaling networks

$$C_f(F) = \max_{P(X)} \{H(X) - H(f(X)|F(X))\} = \max_{P(X)} \{I(f(X);F(X)) + H(X|f(X))\}, \quad (2)$$

where $C_f(F)$ is the computation capacity of the erroneous function $F$ with respect to the error-free function $f$ [4]. In analogy with the communication rate $R(F)$, we will refer to $R_f(F) = H(X) - H(f(X)|F(X))$, for a fixed input distribution $P(X)$, as the *computation rate*. Recall that we have two different input-output network mappings: the error-free correct mapping $f$ and the erroneous incorrect mapping $F$. For modeling and analysis of abnormal signaling networks with some dysfunctional molecules, equation (2) is a more suitable metric than equation (1). This is because it emphasizes the differences between the correct and incorrect outputs of a network, caused by some abnormal conditions and dysfunctional molecules. This is reflected in the new equivocation term $H(f(X)|F(X))$ in the first definition in equation (2), which represents the ambiguity on $f(X)$, the correct network output, given $F(X)$, the incorrect network output. Note that this is different from the traditional equivocation term in equation (1), $H(X|F(X))$ defined in the paragraph after equation (1), which measures the ambiguity on $X$, the network input, given $F(X)$, the incorrect network output. The equivocation model $H(X|F(X))$ is more suitable for communication channels, *where ideally one would like to have the incorrect outputs as close as possible to the inputs*. In contrast, the equivocation model $H(f(X)|F(X))$ is more appropriate for mapping networks and computing systems such as signaling networks, *where ideally it is desired to have the incorrect outputs as close as possible to the correct outputs*.

The second definition in equation (2), developed in Supplementary Material, provides another way to relate computation and communication capacities. It shows that the computation capacity equals the maximum mutual information between correct and incorrect outputs $I(f(X);F(X))$, and the uncertainty on the input given the correct output $H(X|f(X))$. The former is a measure of accuracy of the incorrect output with respect to the correct output, whereas the latter measures the degree of "non-invertibility" of the function $f$, i.e., the degree of the function $f$ "not being one-to-one." Mathematical details and some numerical examples are provided in Supplementary Material.

The summary illustrative table 1 and figure 6 presented at the end of the paper, summarizing our results and findings, further assist with understanding the differences between the computation and communication capacities and their implications for signaling networks.

**A Relation between Computation and Communication Capacities:** An interesting property of the computation capacity in equation (2) is that it is greater than the communication capacity in equation (1), i.e., $C_f(F) > C(F)$ [4], as long as $f$ is not a one-to-one function (no one-to-one correspondence between the elements of domain and co-domain of $f$). This is because the ambiguity of $f(X)$ given $F(X)$ is less than the ambiguity of $X$ given $F(X)$, i.e., $H(f(X)|F(X)) < H(X|F(X))$ by the data processing inequality [6]. Therefore, upon comparing equation (2) with (1), we observe that $C_f(F) > C(F)$, i.e., computation capacity is greater than communication capacity. In signaling networks with lots of redundancies and many cross-linked pathways from inputs to outputs [8], most often the network response function $f$ is not a one-to-one function of the input signals, and therefore the computation capacity is higher than the communication capacity. This is further verified and demonstrated in this paper for two examples of caspase3 and NFκB signaling networks (other networks can be similarly analyzed). Note that only for the special case of $f$ being a one-to-one function, the two capacities become equal.

**Case Study 1) Caspase3 Signaling Network:** Here we calculate the communication and computation capacities of an experimentally-verified signaling network, the caspase3 network (figure 1). Caspase3 is an important molecule and a key regulator of apoptosis. Signaling pathways from the ligands EGF, epidermal growth factor, insulin and TNF, tumor necrosis factor, to caspase3 (figure 1) are extensively characterized and experimentally verified [9]. Based on the experimental results [9], the network output caspase3 is active, when the inputs EGF and insulin are inactive and the input TNF is active. Otherwise, the output is inactive. When a molecule in the network is dysfunctional, one can consider that the activity state of that molecule does not change in response to its regulators [10]. Here we consider the scenario where the dysfunctional molecule remains inactive [10]. This gives rise to a signaling network that can be considered as an erroneous computing machine. The input-output functional relationship $F$ of the abnormal network can be different from $f$ of the normal network, and depends on which molecule is dysfunctional, as shown in Supplementary Material, Section A. We observe that when AKT or EGFR or MEKK1ASK1 is dysfunctional in the network, the input-output abnormal mapping function $F$ is different from $f$ of the normal network; whereas $F$ is the same as $f$, when other molecules in the network are dysfunctional.

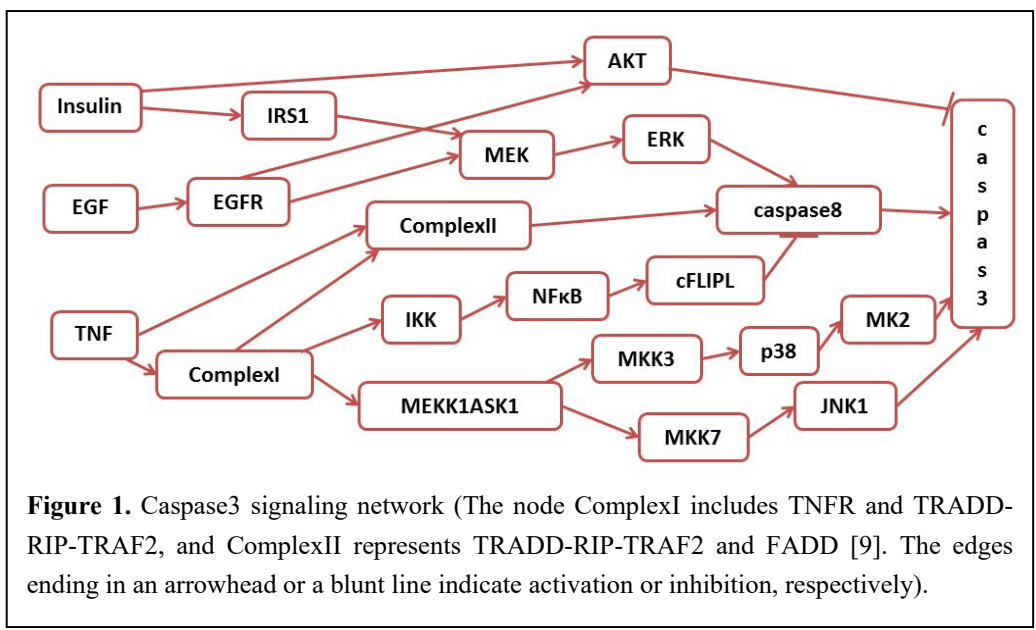

**Figure 1.** Caspase3 signaling network (The node ComplexI includes TNFR and TRADD-RIP-TRAF2, and ComplexII represents TRADD-RIP-TRAF2 and FADD [9]. The edges ending in an arrowhead or a blunt line indicate activation or inhibition, respectively).

In what follows, to quantify the amount of impact of each dysfunctional molecule on communication and computation capacities of the caspase3 signaling network, we consider a model where each single molecule can be dysfunctional with a probability $p$, such that $0 \leq p \leq 1$.

**Communication Capacity $C(F)$ of the Caspase3 Signaling Network:** Using the network transition probability matrices of the caspase3 network, equations (2)-(5) of [11], and based on equation (1) in this paper, the communication capacity of the caspase3 network is plotted in figure 2, when one of its molecules is dysfunctional (see Supplementary Material for the method). Interestingly, the communication capacity metric overlooks the differences between the molecules and classifies them into two groups. It appears that the communication capacity provides less information about the network and its abnormal behavior, when it contains dysfunctional molecules. This is in contrast to the computation capacity, as discussed next.

**Computation Capacity $C_f(F)$ of the Caspase3 Signaling Network:** Using the network transition probability matrices of the caspase3 network, equations (2)-(5) of [11], and based on equation (2) in this paper, the computation capacity of the caspase3 network with respect to the error-free output $f$ is plotted in figure 3, when one of its molecules is dysfunctional (see Supplementary Material for the method). Comparing with the communication capacity in figure 2, we notice two remarkable points:

(i) The computation capacity magnitude is larger than the communication capacity. This is because the input-output mapping function $f$ of the caspase3 network (Supplementary Material, equation (s1)) is not a bijective function, i.e., is not a one-to-one correspondence. This makes the computation capacity larger than the communication capacity, as stated earlier in the paper. Intuitively, this indicates that the number of input values that can be correctly computed on is larger than the number of input values that can be correctly recovered given the output of the network.

(ii) The computation capacity classifies the network molecules into four groups. This indicates that the computation capacity can have more predictive power than the communication capacity, which identifies only two groups of molecules. In other words, the computation capacity can recognize the roles and functions of different molecules in the network more precisely than the communication capacity.

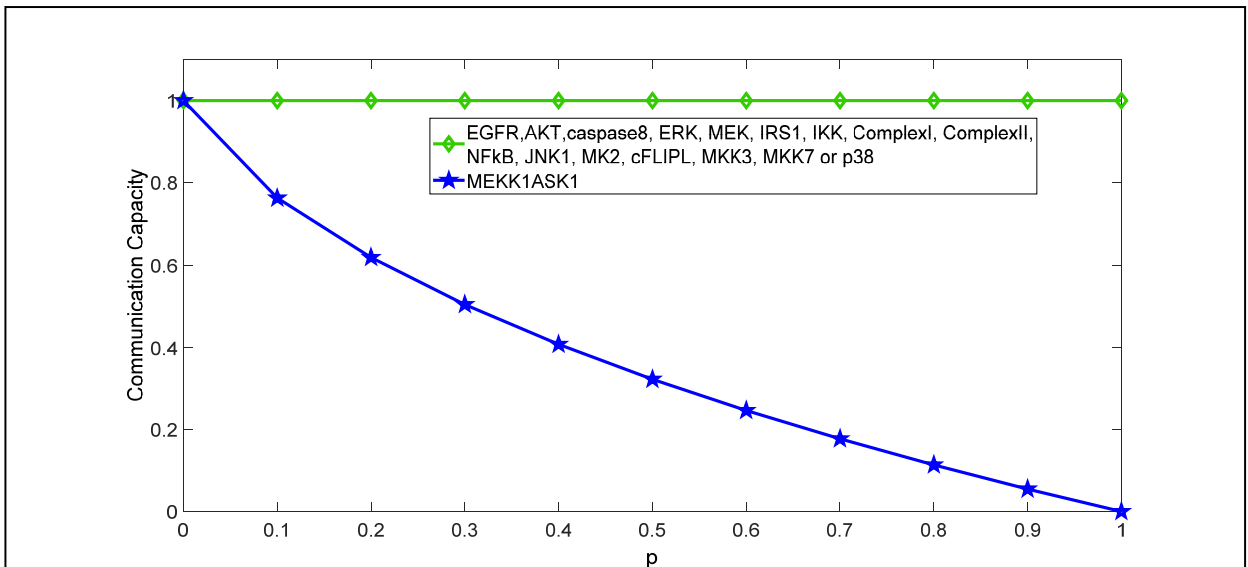

**Figure 2.** Communication capacity in equation (1) versus the dysfunction probability $p$ for each molecule in the caspase3 network.

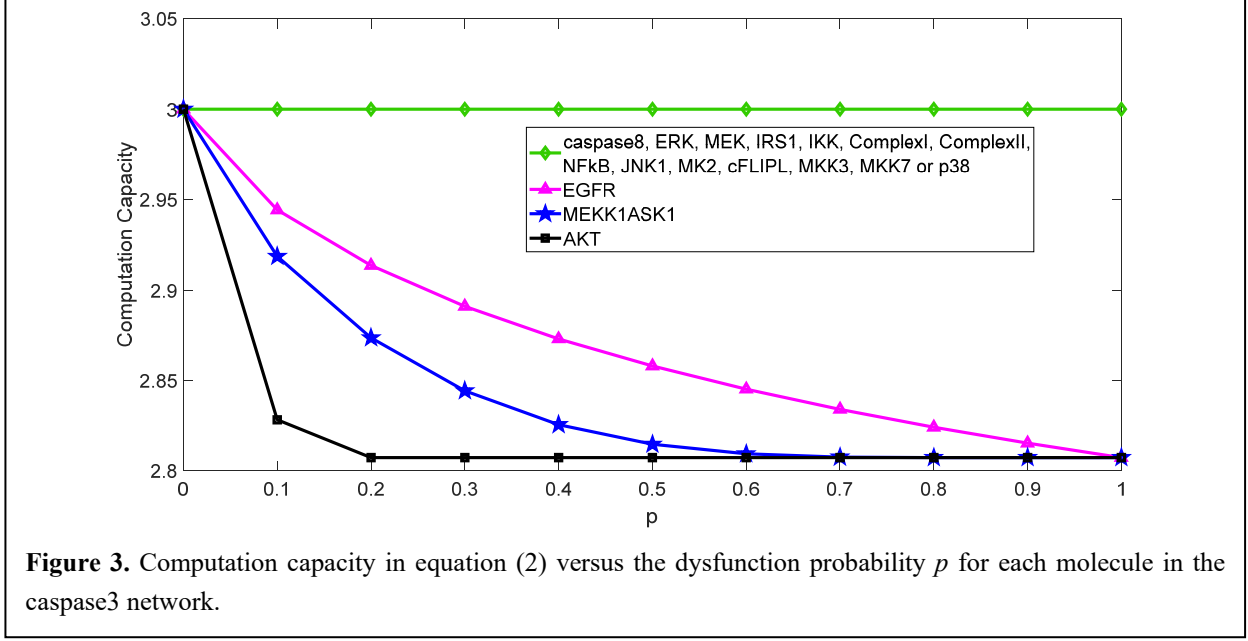

**Figure 3.** Computation capacity in equation (2) versus the dysfunction probability $p$ for each molecule in the caspase3 network.

**Case Study 2) NFκB Signaling Network:** Now we study some communication and computation characteristics of a network that has feedback and hence defines an input-output mapping $f$ with memory. Consider the network in figure 4, where tumor necrosis factor (TNF) and nuclear factor κB (NFκB) are input and output molecules, respectively. The molecule A20 has an inhibitory feedback effect, whereas TRC stands for TNF receptor complex [5]. A comprehensive stochastic differential equation model is developed in [12] and its accuracy is extensively verified via experimental data. It is well known that the activity of NFκB first increases with TNF, but the upregulation of A20 by NFκB inhibits TRC, which in turn decreases the activity of NFκB. Consider that the inactive and active states of a molecule are represented by 0 and 1, respectively (see [10,13,14,15] for an overview and examples of this modeling approach in systems biology). In Supplementary Material, Sections C, D and E, we first present activity models for the network output in term of the input activity, under normal (wild type) and abnormal (A20-deficient) conditions, and show how the input-output models are corroborated by biological data. Afterwards, we present a system formulation and discuss its biological relevance in Section F of Supplementary Material. The system formulation allows to calculate and compare communication and computation rates and capacities of the NFκB network.

We consider the network to be abnormal, when it has A20 deficiency. It is demonstrated that an A20-deficient mouse develops severe inflammation and dies prematurely [16]. This is because cells with this deficiency cannot stop the NFκB response caused by TNF, as is evident in figure 4, when there is no feedback. It is also known that the dysfunction of A20 is involved in a number of diseases such as multiple sclerosis, lupus, rheumatoid arthritis, etc. [17]. To model A20 deficiency in the network, we consider A20 as a molecule which has a chance to be dysfunctional with a probability *p*. More specifically, consider that the probability of A20 to remain 0, inactive, regardless of the signal from NFκB is *p*. This model is consistent with the fact that A20 is inactive in several hematological malignancies [18], and also as shown below, it allows to calculate and compare communication and computation rates and capacities of the NFκB network, when A20 is dysfunctional.

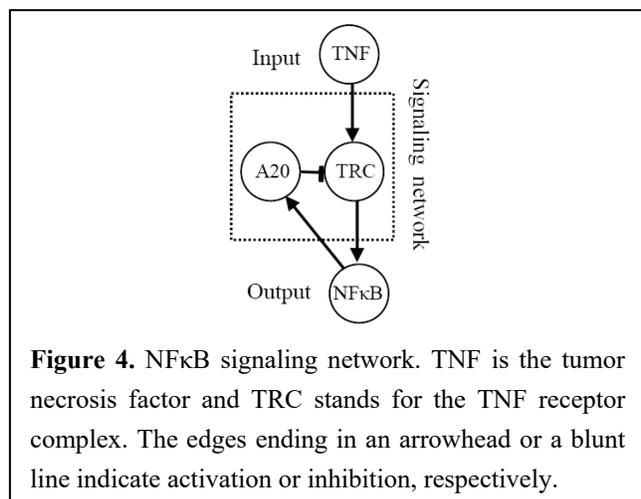

**Figure 4.** NFκB signaling network. TNF is the tumor necrosis factor and TRC stands for the TNF receptor complex. The edges ending in an arrowhead or a blunt line indicate activation or inhibition, respectively.

**Communication Rate of the NFκB Signaling Network:** Due to the feedback in the network, the network has a memory such that its output (NFκB) activity state depends on the present and past input (TNF) activity states (see Section G of Supplementary Material). Calculation of the communication capacity in the presence of memory requires to optimize over the distribution of sequences of inputs. In order to gain some insight into the communication capacity of systems with memory, we first evaluate the communication rate, i.e., the mutual information between two successive values of the input and the corresponding output values for a fixed uniform distribution of the input.

To elaborate, let $Y_1$ and $Y_2$ represent the activity states of NFκB at two consecutive time instants $t=1$ and $t=2$, respectively, i.e., $Y_1 = \text{NFκB}(t=1)$ and $Y_2 = \text{NFκB}(t=2)$. Note that $t=1$ and $t=2$ represent early and late signaling events, respectively. Similarly, we have $X_1 = \text{TNF}(t=1)$ and $X_2 = \text{TNF}(t=2)$. Clearly $X$ and $Y$ variables refer to the network input and output in figure 4, respectively. The communication rate evaluated as the mutual information between the network input sequence $(X_1, X_2)$ and the network output sequence $(Y_1, Y_2)$ can be shown to be (see Section G of Supplementary Material for the method)

$$\begin{aligned} R(F) &= I(X_1, X_2; Y_1, Y_2) = H(X_1, X_2) - H(X_1, X_2 | Y_1, Y_2) \\ &= 2 - 0.25(2-p)\log_2(2-p) + 0.25(1-p)\log_2(1-p), \end{aligned} \quad (3)$$

where $\log_2$ is logarithm to the base 2 and *p* is the probability of A20 to be dysfunctional. As seen in figure 5, the communication rate, i.e., the mutual information, increases with *p*. This means the network output in figure 4 has more information about the input, when A20 becomes more dysfunctional. In other words, as A20 becomes more dysfunctional, the network behaves closer to a linear pathway with no feedback, and the input activity state can be determined with less ambiguity from the output activity state.

**Computation Rate of the NFκB Signaling Network:** Using the notation introduced earlier, i.e., $Y_1 = \text{NF}\kappa\text{B}(t=1)$, $Y_2 = \text{NF}\kappa\text{B}(t=2)$, $X_1 = \text{TNF}(t=1)$ and $X_2 = \text{TNF}(t=2)$, the computation rate of the network with respect to the error-free output $f$ can be shown to be (see Section G of Supplementary Material for the method)

$$R_f(F) = H(X_1, X_2) - H(f(X_1, X_2) | Y_1, Y_2) = 2. \tag{4}$$

As seen in figure 5, the computation rate for the NFκB network is greater than its communication rate. Moreover, its constant value implies that regardless of $p$, correct outputs can be inferred from erroneous outputs (Supplementary Material, Section G).

**Communication and Computation Capacities of the NFκB Signaling Network:** Let $C(F)$ represent the communication capacity of the NFκB network. It can be shown that (see Section H of Supplementary Material for the methods)

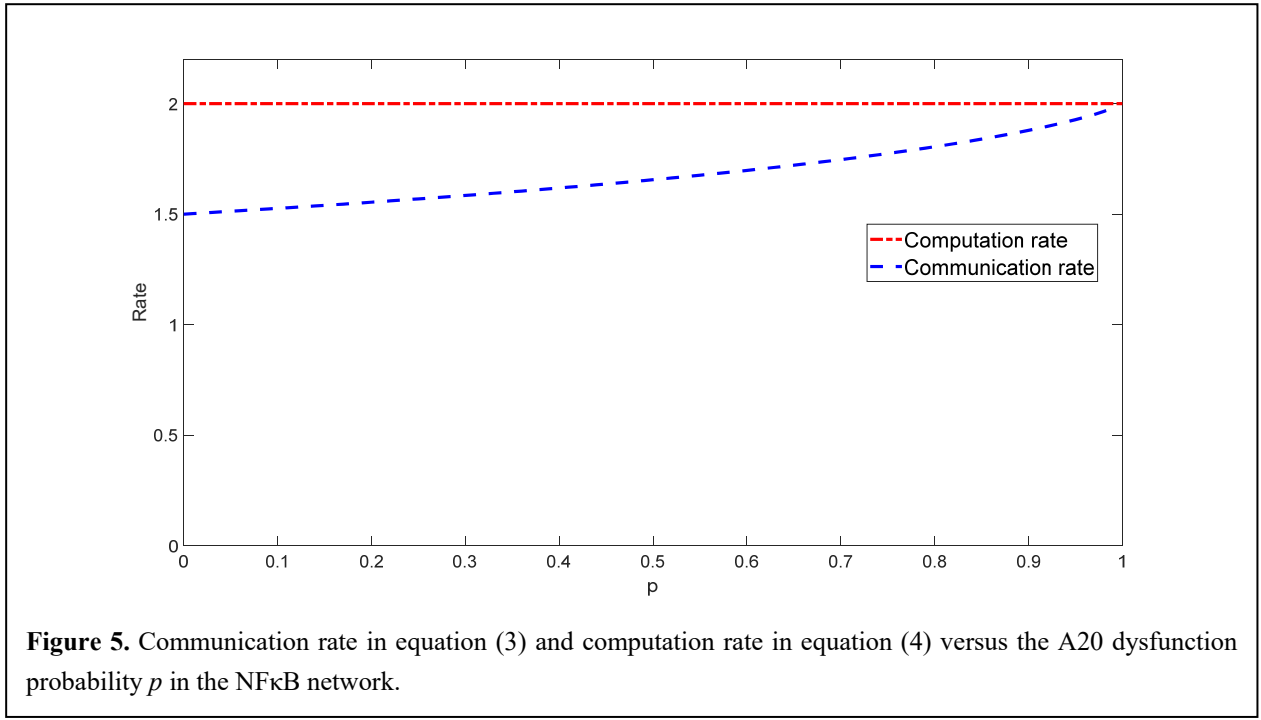

**Figure 5.** Communication rate in equation (3) and computation rate in equation (4) versus the A20 dysfunction probability $p$ in the NFκB network.

$$C(F) = \begin{cases} \log_2\left(\dfrac{1+\sqrt{5}}{2}\right) \approx 0.7, & 0 \leq p < 1, \\ 1, & p = 1. \end{cases} \quad (5)$$

Interestingly, the term $(1+\sqrt{5})/2$ is the well-known golden ratio [19]. Another noteworthy observation is that the calculated communication capacity of about 0.7 bits for the TNF-NFκB system is based on early and late responses of NFκB to TNF. It falls between the experimentally-determined individual maximum mutual information of about 0.9 and 0.6 bits for the same system, based on early and late NFκB responses, respectively [5].

On the other hand, let $C_f(F)$ stand for the computation capacity of the NFκB network with respect to the error-free output $f$. It can be shown that (see Section I of Supplementary Material for the methods)

$$C_f(F) = 1, \ 0 \leq p \leq 1. \quad (6)$$

Note that since the function $f$ here is not one-to-one (see the second column of table S2), the computation capacity, 1 in equation (6), is higher than the communication capacity, 0.7 in equation (5). Only for the special case of $p=1$, A20 being completely dysfunctional with unit probability, the NFκB network (figure 4) becomes a linear pathway in the absence of A20 feedback. This makes $f$ a one-to-one function that results in equal capacities for the special case of $p=1$.

**On the Computation Capacity Being Higher than the Communication Capacity:** We generally discussed earlier in the paper that signaling networks with lots of redundancies and many cross-linked pathways from inputs to outputs do not possess one-to-one mapping functions, and therefore their computation capacities are proved to be higher than their communication capacities. We also verified this specifically for two signaling networks. Interestingly, it has been a surprise to researchers that information (communication) capacities of signaling networks appear to be smaller than what is typically anticipated [1,20]. In this regard, the computation capacity concept introduced here is a new metric that can possibly shed some light on this controversy, and can perhaps reveal some unknown capabilities and characteristics of signaling networks.

**An Illustrative Comparative Summary of Computation and Communication Capacities of Signaling Networks:** The computation capacity is introduced in this paper for normal and abnormal networks, i.e., networks with dysfunctional molecules, whereas the communication capacity of normal and abnormal networks was previously studied in [2]. Some other studies have investigated the communication capacity of signaling networks [5,21,22,23] and genetic systems [24].

To better understand the differences between the computation and communication capacity metrics, we provide an illustrative schematic in figure 6. The figure depicts a signaling network with some dysfunctional molecules, the erroneous input-output mapping function $F$, the input signal $X$ and the erroneous output response $F(X)$. As graphically shown in figure 6, the communication capacity $C(F)$ considers only $X$ and $F(X)$. In contrast, the computation capacity $C_f(F)$ additionally takes into account the desired error-free input-output mapping function $f$. In other words, the computation capacity $C_f(F)$ considers $X$, $F(X)$ and $f(X)$ (figure 6). The transformation $\Phi$ (figure 6) is a mapping from $f$ to $F$, and models the changeover from the normal response $f(X)$ to the abnormal response $F(X)$. In other words, $\Phi$ represents computation/signaling errors caused by dysfunctional components and molecules. Note that the abnormal and normal states in figure 6 are symbolically shown by an irregular shape and a rectangle, respectively. Overall, this figure elucidates the concepts behind the communication and computation capacities and their differences.

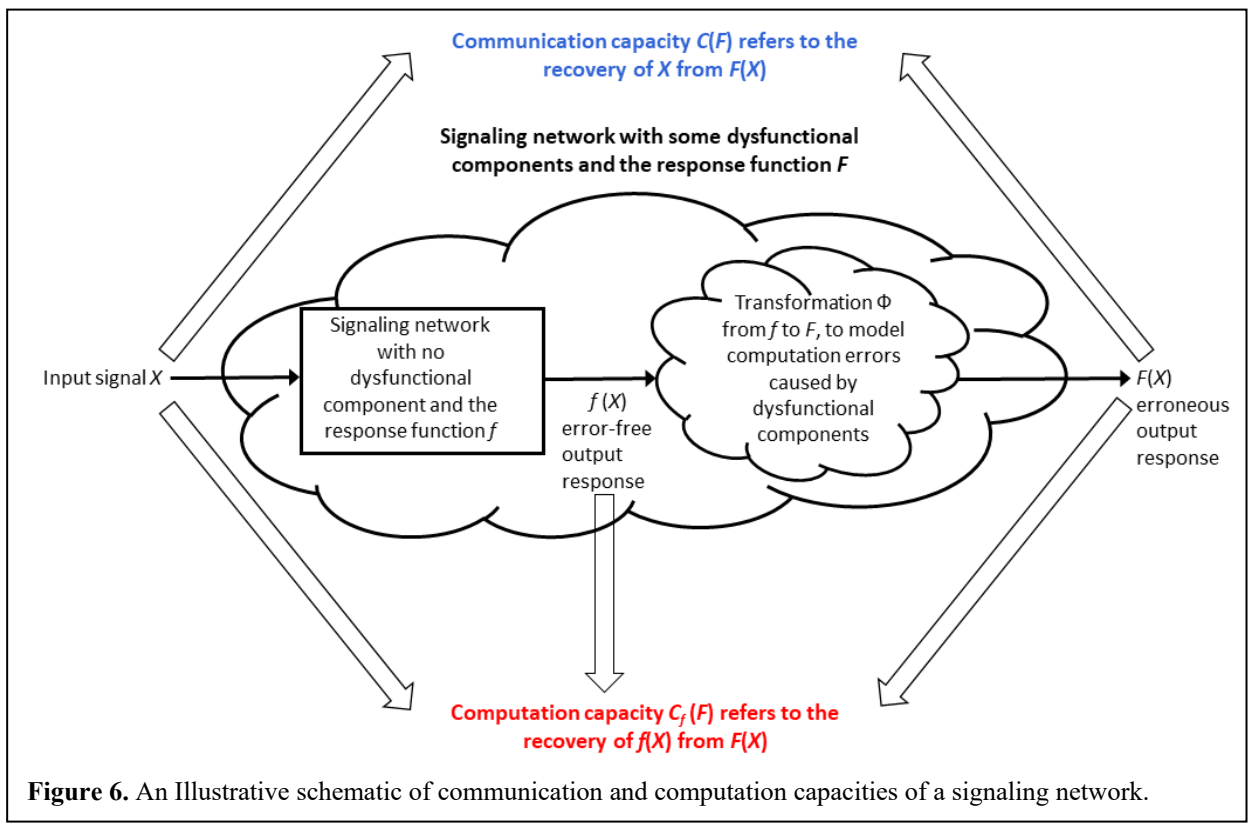

**Figure 6.** An Illustrative schematic of communication and computation capacities of a signaling network.

Additionally, a summary of communication and computation capacity-related metrics and definitions and results that clarifies their differences is presented in table 1. The corresponding mathematical details and derivations and other numerical examples for communication and computation capacities are presented in Supplementary Material, Sections J-L.

| $X$: input signal $f(X)$: error-free network output $F(X)$: erroneous network output $H$: entropy | Equivocation | Capacity | $2^{Capacity}$ |
|---|---|---|---|
| Communication metrics | $H(X\|F(X))$ | $C(F) = \max$ of $H(X) - H(X\|F(X))$ | Max number of $X$ values such that $X$ can be correctly recovered from $F(X)$, i.e., max number of decodable or distinguishable inputs. |
| Computation metrics | $H(f(X)\|F(X))$ | $C_f(F) = \max$ of $H(X) - H(f(X)\|F(X))$ | Max number of $X$ values such that $f(X)$ can be correctly recovered from $F(X)$, i.e., max number of computable inputs. |

**Table 1.** Summary of communication and computation capacity-related metrics and definitions and results.

**On Signaling Network Models for Calculating Computation and Communication Capacities:** In the stochastic network models considered in this paper, activity level of each molecule is a continuous-valued number between 0 and 1, indicating the probability of the molecule to be active [15] (For an overview and examples of this modeling approach in systems biology, interested readers can refer to [10,13,14,15]). To model the presence of feedback in the NFκB network, a time-varying model is developed in this paper. Additionally, we have used the experimentally verified stochastic differential equation model of [12], also used in [25], to generate data. We have used the data to demonstrate the biological relevance of our developed model, which is suitable for calculating and comparing communication and computation capacities of the NFκB network, under similar conditions. Extension of the computation capacity concept to concentration-type models such as differential equation-based models is a possible next step. In this context, care should be taken when defining entropies for such models. Presence of memory and time variations in a system under study make capacity definitions and calculations particularly difficult, due to the need to optimize over the input distribution. If not feasible to calculate the computation capacity for such scenarios, still it is helpful to calculate the computation rate instead, to gain some insights.

**Conclusions:** Cell signaling networks can be envisioned as computing systems that compute the outputs in response to the inputs. The system inputs can be considered to be ligands which upon binding to their receptors on the cell membrane, create chains of interactions through some intermediate signaling molecules. The system outputs are some target proteins such as transcription factors. Due to the presence of dysfunctional molecules in a signaling network, it may behave abnormally, i.e., may compute the network outputs incorrectly. In this paper, a new fundamental characteristic of signaling networks, i.e., the computation capacity, is introduced and investigated. Our results on caspase3 and NFκB networks indicate that their computation capacities are higher than their communication capacities. Additionally, it is shown in the paper that in general, the network computation capacity is higher than the network communication capacity, as long as the network response function is not a one-to-one function of the input signals. One biological implication of this finding is that signaling networks may have more capabilities than what we presently know. Overall, this study and its findings are anticipated to advance our understanding of some fundamental characteristics of cell signaling networks.

# Supplementary Material

**Computation capacities of a broad class of signaling networks are higher than their communication capacities**

Iman Habibi[1], Effat S. Emamian[2], Osvaldo Simeone[3] and Ali Abdi[4,5] *

[1] Iman Habibi
Center for Wireless Information Processing, Department of Electrical and Computer Engineering
New Jersey Institute of Technology, 323 King Blvd, Newark, NJ 07102, USA
Email: ih26@njit.edu

[2] Effat S. Emamian, MD
Advanced Technologies for Novel Therapeutics
Enterprise Development Center
New Jersey Institute of Technology, 211 Warren St., Newark, NJ 07103, USA
Email: emame@atnt-usa.com

[3] Osvaldo Simeone, PhD
King's Centre for Learning & Information Processing (kclip)
Center for Telecommunications Research
Department of Engineering
King's College London, London, WC2R 2LS, England, UK
Email: osvaldo.simeone@kcl.ac.uk

Ali Abdi, PhD
[4] Center for Wireless Information Processing, Department of Electrical and Computer Engineering
[5] Department of Biological Sciences
New Jersey Institute of Technology, 323 King Blvd, Newark, NJ 07102, USA
Email: ali.abdi@njit.edu

* Corresponding Author: Ali Abdi (ali.abdi@njit.edu)

This Supplementary Material includes the following sections:

*Caspase3 network modeling and capacity analysis:*

A. Method to determine the input-output functional relationships $f$ and $F$ for the normal and abnormal caspase3 networks, respectively

B. Method for calculating the communication and computation capacities of the caspase3 signaling network

*NFκB network modeling:*

C. Activity models for the NFκB network output in term of the input TNF activity

D. Method for obtaining the input-output activity equation of the normal NFκB network

E. Method for obtaining the input-output activity equation of the abnormal NFκB network

F. System formulation suitable for calculating communication and computation rates and capacities of the NFκB network

*NFκB network capacity analysis:*

G. Method for calculating communication and computation rates of the NFκB network

H. Method for calculating communication capacity of the NFκB network

I. Method for calculating computation capacity of the NFκB network

*Mathematical details and numerical examples for communication and computation capacities:*

J. Capacity definitions and formulas

K. An example of a malfunctioning system and its communication and computation capacities

L. Communication and computation coding theorems for malfunctioning systems

**A. Method to determine the input-output functional relationships *f* and *F* for the normal and abnormal caspase3 networks, respectively.** Consider that the inactive and active states of a molecule are represented by 0 and 1, respectively (see [Wang12], [Saad13], [Heli08], [Abdi08] and references therein for an overview and examples of this modeling approach in systems biology). On the other hand, based on the experimental results of [Jane06], the network output caspase3 in figure 1 is active, when the inputs EGF and insulin are inactive and the input TNF is active. Otherwise, the output is inactive. Consistent with the experimental findings of [Jane06], the input-output functional relationship *f* of the normal network, i.e., when all the molecules in the network are functioning properly and there is no dysfunctional molecule, can be written as

$$\text{caspase3} = f(\text{EGF}, \text{insulin}, \text{TNF}) = \begin{cases} 1, & (\text{EGF}, \text{insulin}, \text{TNF}) = (0,0,1), \\ 0, & \text{otherwise}. \end{cases} \quad (s1)$$

This represents an error-free computing machine which computes a value for the machine output caspase3, based on the values of the machine inputs EGF, insulin and TNF. Regulatory equations [Abdi08] for the caspase3 network (figure 1) which reproduce the experimentally-verified normal network response function *f* introduced in equation (s1) are listed in table S0. Each regulatory equation specifies how the activity state of each molecule is determined by the activity states of its input signals, using the logic operations ', + and ×, which stand for NOT, OR and AND, respectively.

To model signaling errors, we consider a molecule to be dysfunctional, if it remains inactive, stuck at 0, regardless of the signals from its regulators [Abdi08]. For example, when MEKK1ASK1 is stuck at 0, regardless of its input signals, then using the equations in table S0, it can be verified that the output remains inactive all the times. This results in the abnormal network response function *F* presented in equation (s4). Note that *F* in equation (s4) is different from the normal network response function *f* in equation (s1). To characterize all abnormal caspase3 network responses, the following input-output functional relationships are similarly derived for *F*, depending on which molecule is dysfunctional (these results agree with Table S2 of [Habi14a] as well)

| Molecule Name | Regulatory Equation |
|---|---|
| AKT | AKT=EGFR+Insulin |
| Caspase3 | Caspase3=AKT'×(Caspase8+JNK1+MK2) |
| Caspase8 | Caspase8=cFLIP$_L$'×(ComplexII+ERK) |
| cFLIP$_L$ | cFLIP$_L$=NFκB |
| ComplexI | ComplexI=TNF |
| ComplexII | ComplexII=TNF+ComplexI |
| EGFR | EGFR=EGF |
| ERK | ERK=MEK |
| IKK | IKK=ComplexI |
| IRS1 | IRS1=Insulin |
| JNK1 | JNK1=MKK7 |
| MEK | MEK=EGFR+IRS1 |
| MEKK1ASK1 | MEKK1ASK1=ComplexI |
| MK2 | MK2=p38 |
| MKK3 | MKK3=MEKK1ASK1 |
| MKK7 | MKK7=MEKK1ASK1 |
| NFκB | NFκB=IKK |
| p38 | p38=MKK3 |

**Table S0.** Regulatory equations [Abdi08] for the caspase3 network (figure 1).

AKT is dysfunctional

$$\text{caspase3} = F(\text{EGF}, \text{insulin}, \text{TNF}) = \begin{cases} 0, & (\text{EGF}, \text{insulin}, \text{TNF}) = (0,0,0), \\ 1, & \text{otherwise.} \end{cases} \quad (s2)$$

EGFR is dysfunctional

$$\text{caspase3} = F(\text{EGF}, \text{insulin}, \text{TNF}) = \begin{cases} 1, & (\text{EGF}, \text{insulin}, \text{TNF}) = (0,0,1) \text{ or } (1,0,1), \\ 0, & \text{otherwise.} \end{cases} \quad (s3)$$

MEKK1ASK1 is dysfunctional

$$\text{caspase3} = F(\text{EGF}, \text{insulin}, \text{TNF}) = 0. \quad (s4)$$

Another molecule is dysfunctional

$$\text{caspase3} = F(\text{EGF}, \text{insulin}, \text{TNF}) = \begin{cases} 1, & (\text{EGF}, \text{insulin}, \text{TNF}) = (0,0,1), \\ 0, & \text{otherwise.} \end{cases} \quad (s5)$$

Note that in equations (s2)-(s4) we have $F \neq f$, whereas $F = f$ in equation (s5).

**B. Method for calculating the communication and computation capacities of the caspase3 signaling network.** To calculate the capacities of this network (figure 1), one needs its network transition probability matrices. Depending on which molecule is dysfunctional, we have used the network transition probability matrices provided in equations (2)-(5) of [Habi14b]. The Arimoto algorithm [Arim72] is used afterwards, to numerically calculate the communication and computation capacities graphed in figure 2 and figure 3, respectively.

**C. Activity models for the NFκB network output in term of the input TNF activity:** We have derived the following equation for the activity of NFκB in terms of the TNF activity (see Section D of Supplementary Material for the method)

$$P(\text{NF}\kappa\text{B} = 1) = q/(1+q), \qquad (s6)$$

where $P(.)$ stands for probability and $q = P(\text{TNF} = 1)$ denotes the TNF activity. Note that the probability of a molecule to be active can be considered as the activity of the molecule [Heli08]. In figure S1, we observe that equation (s6) agrees with the average activity data [Lipn07] [Tay10], obtained from 200 single cell results. In the figure, equation (s6) is normalized such that it matches the average activity data at $q = 1$.

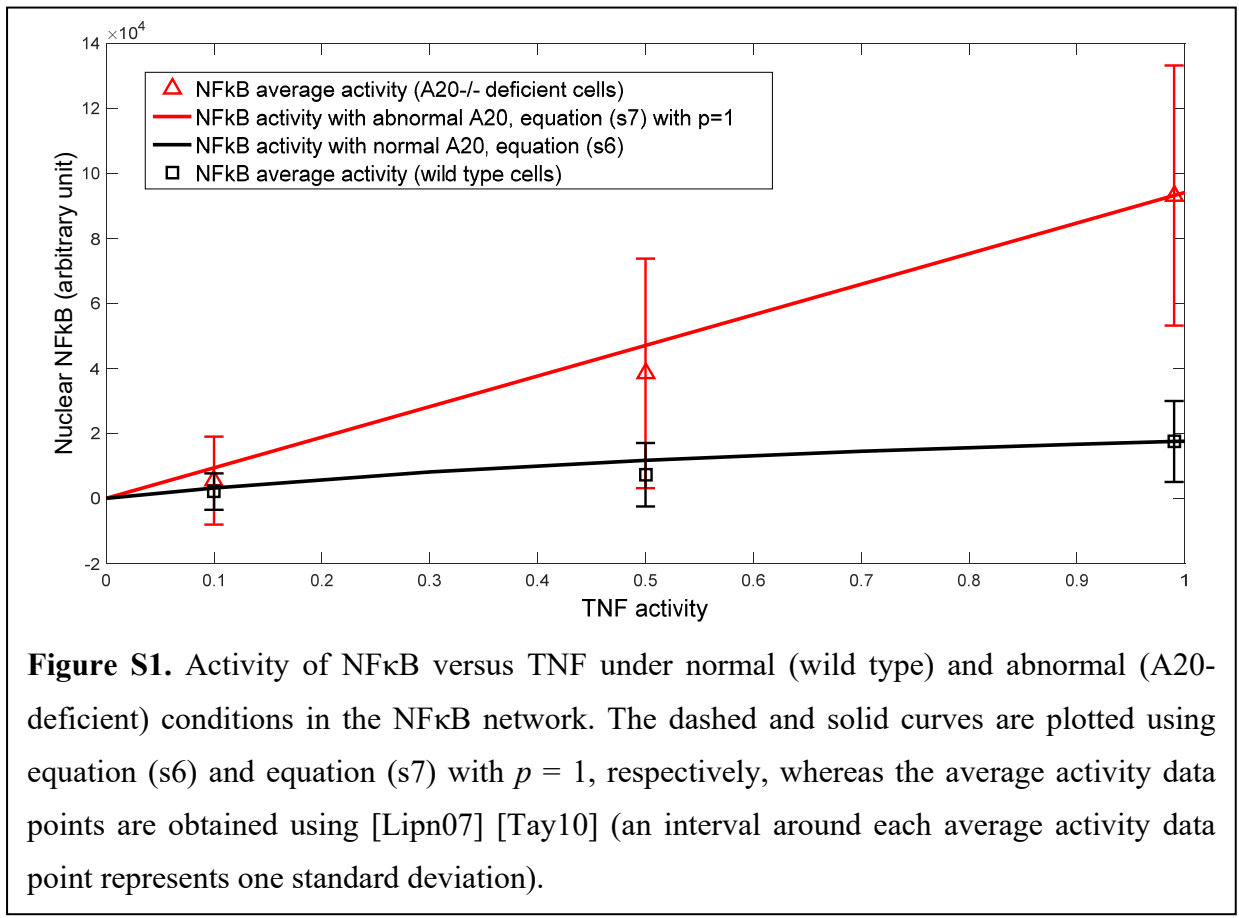

**Figure S1.** Activity of NFκB versus TNF under normal (wild type) and abnormal (A20-deficient) conditions in the NFκB network. The dashed and solid curves are plotted using equation (s6) and equation (s7) with $p = 1$, respectively, whereas the average activity data points are obtained using [Lipn07] [Tay10] (an interval around each average activity data point represents one standard deviation).

Now we consider an abnormal network due to its A20 deficiency. It is demonstrated that an A20-deficient mouse develops severe inflammation and dies prematurely [Lee00]. This is because cells with this deficiency cannot stop the NFκB response caused by TNF, as is evident in figure 4, when there is no feedback. It is also known that the dysfunction of A20 is involved in a number of diseases such as multiple sclerosis, lupus, rheumatoid arthritis, etc. [Vers10]. To model A20 deficiency in the network, we consider A20 as a molecule which has a chance to be dysfunctional with a probability $p$. More specifically, consider that the probability of A20 to remain 0, inactive, regardless of the signal from NFκB is $p$. This model is consistent with the fact that A20 is inactive in several hematological malignancies [Hymo10]. This modeling approach has allowed us to derive the following equation for the activity of NFκB in terms of the TNF activity, when A20 is dysfunctional (see Section E of Supplementary Material for the method)

$$P(\text{NF}\kappa\text{B} = 1) = q(1 + pq)/(1 + q). \qquad (s7)$$

For $p = 0$, normal A20, equation (s7) reduces to (s6), as expected. When A20 is completely dysfunctional, i.e., $p = 1$, equation (s7) results in $P(\text{NF}\kappa\text{B} = 1) = q$, which is a linear relationship. This is consistent with the linear network structure in figure 4 when there is no feedback, and also agrees with the biology of this network, i.e., persistent NFκB signaling stimulated by TNF in A20-deficient cells [Hymo10].

Comparison of equation (s7) with the A20-inactivated NFκB levels [Lipn07] [Tay10] in figure S1 demonstrates that the activity model developed for the abnormal network is biologically relevant (A20-inactivated data is obtained by setting the parameter AB to zero [Lipn07] [Tay10], which induces zero A20 mRNA synthesis). Moreover, the difference between the activity levels of NFκB under normal and abnormal conditions indicates transition from the normal to the abnormal state. This difference is caused by the dysfunction of A20.

**D. Method for obtaining the input-output activity equation of the normal NFκB network:**

Here we derive an expression for the activity of NFκB in terms of the TNF activity, using the biological interactions in the NFκB network (figure 4). Note that the probability of a molecule to be active can be considered as the activity of the molecule [Heli08]. Therefore, activities of NFκB and TNF can be represented by $P(\text{NFκB}=1)$ and $q = P(\text{TNF}=1)$, respectively. Using the total probability theorem [Papo91] we have

$$P(\text{NFκB}=1) = P(\text{NFκB}=1\,|\,\text{TNF}=0)P(\text{TNF}=0) + P(\text{NFκB}=1\,|\,\text{TNF}=1)P(\text{TNF}=1), \quad (s8)$$

where the notation | stands for conditional probability. When TNF is inactive, NFκB is also inactive, which means $P(\text{NFκB}=1\,|\,\text{TNF}=0) = 0$. This simplifies equation (s8) to

$$P(\text{NFκB}=1) = P(\text{NFκB}=1\,|\,\text{TNF}=1)\,q. \quad (s9)$$

Now we relate the activity of NFκB to the activity of A20, by re-writing equation (s9) using the total probability theorem, in terms of different activity levels of A20

$$\begin{aligned}P(\text{NFκB}=1) = [&P(\text{NFκB}=1\,|\,\text{TNF}=1, \text{A20}=0)P(\text{A20}=0) \\ +\ &P(\text{NFκB}=1\,|\,\text{TNF}=1, \text{A20}=1)\,P(\text{A20}=1)]\,q.\end{aligned} \quad (s10)$$

When TNF is active and A20 is inactive (no inhibition in figure 4), NFκB also becomes active, which means $P(\text{NFκB}=1\,|\,\text{TNF}=1, \text{A20}=0) = 1$. On the other hand, when A20 is active, it inhibits TRC (figure 4) which in turn makes NFκB inactive [Cheo11], i.e., $P(\text{NFκB}=1\,|\,\text{TNF}=1, \text{A20}=1) = 0$. By inserting these results in (s10) we obtain

$$P(\text{NFκB}=1) = P(\text{A20}=0)\,q. \quad (s11)$$

Since NFκB regulates the activity of A20 (figure 4), we can say $P(\text{NFκB}=0) = P(\text{A20}=0)$, which changes equation (s11) to $P(\text{NFκB}=1) = P(\text{NFκB}=0)\,q$. By replacing $P(\text{NFκB}=0)$ with $1 - P(\text{NFκB}=1)$ and solving for $P(\text{NFκB}=1)$, we obtain equation (s6) for the activity of NFκB.

**E. Method for obtaining the input-output activity equation of the abnormal NFκB network:**

When A20 is dysfunctional, the NFκB network diverges from its normal functions. Using the total probability theorem [Papo91], $P(A20=0)$ can be written as

$$P(A20=0) = P(A20=0 \mid A20 \text{ is dysfunctional}) P(A20 \text{ is dysfunctional}) \\ + P(A20=0 \mid A20 \text{ is not dysfunctional}) P(A20 \text{ is not dysfunctional}). \quad (s12)$$

Let $p = P(A20 \text{ is dysfunctional})$ denote the probability of A20 being dysfunctional. Also, to model signaling errors and abnormalities, we consider a molecule to be dysfunctional, if it remains inactive (stuck at 0), regardless of the signals from its regulators [Abdi08]. This means $P(A20=0 \mid A20 \text{ is dysfunctional}) = 1$. Therefore, equation (s12) can be written as

$$P(A20=0) = p + P(A20=0 \mid A20 \text{ is not dysfunctional})(1-p). \quad (s13)$$

Since NFκB regulates the activity of A20 (figure 4) when A20 is not dysfunctional, we can say $P(A20=0 \mid A20 \text{ is not dysfunctional}) = P(NF\kappa B=0 \mid A20 \text{ is not dysfunctional})$. Moreover, $P(NF\kappa B=0 \mid A20 \text{ is not dysfunctional}) = 1 - P(NF\kappa B=1 \mid A20 \text{ is not dysfunctional}) = 1-[q/(1+q)]$, where the last expression comes from (s6). By applying $1-[q/(1+q)]$ to (s13) and substituting the resulting expression in (s11), we obtain equation (s7) for the activity of NFκB, when A20 is dysfunctional.

**F. System formulation suitable for calculating communication and computation rates and capacities of the NFκB network:** It is well known that activity of NFκB is first controlled by TNF, but activation of A20 by NFκB inhibits TRC, which in turn decreases the activity of NFκB. To model this behavior and based on figure 4, one can write $NF\kappa B(t) = TRC(t)$ and $A20(t) = NF\kappa B(t)$, which show the activity states of some molecules at time $t$. On the other hand, TRC activity depends on the activity of TNF and the feedback inhibitor A20, i.e., $TRC(t) = TNF(t) \times \overline{A20}(t-1)$, where the bar indicates logical negation and × stands for logical and. By combining these equations we obtain the output activity equation $NF\kappa B(t) = TNF(t) \times \overline{NF\kappa B}(t-1)$, which shows due to the

feedback, activity of NFκB depends on its past activity also. Using this equation, possible activity states of NFκB are shown in table S1, in the column in the middle.

Now we consider an abnormal network due to its A20 deficiency. It is demonstrated that an A20-deficient mouse develops severe inflammation and dies prematurely [Lee00]. This is because cells with this deficiency cannot stop the NFκB response caused by TNF, as is evident in figure 4, when there is no feedback. It is also known that the dysfunction of A20 is involved in a number of diseases such as multiple sclerosis, lupus, rheumatoid arthritis, etc. [Vers10]. To model A20 deficiency in the network, we consider A20 as a dysfunctional molecule such that its activity state remains at 0, i.e., inactive, regardless of the signal from NFκB. This model is consistent with the fact that A20 is inactive in several hematological malignancies [Hymo10]. When A20 is dysfunctional, there is no feedback in the network and the network becomes a linear pathway (figure 4). This means $TRC(t) = TNF(t)$ and $NFκB(t) = TRC(t)$, which eventually results in the output activity equation $NFκB(t) = TNF(t)$, when A20 is dysfunctional. Using this equation, possible activity states of NFκB are shown in table S1, in the last column.

*Comparison with biological activity data:* In [Lipn07] experimental data are provided for the NFκB activity, under persistent and pulse stimulations by TNF. In what follows we show predictions of the system formulation presented in this section are corroborated by biological data. First consider persistent TNF stimulation, which using our notation means $TNF(t) = 1$ all the time. Assuming no initial NFκB activity, $NFκB(t = 0) = 0$, the "1, 0" row of table S1 results in NFκB($t=1$) $=1$, which means NFκB becomes active. At the next time point $t=2$, however, NFκB becomes inactive due to the inhibitory A20 feedback, as listed in the "1, 1" row of table S1, therefore, NFκB($t=2$) $=0$. As long as TNF remains active, this NFκB activity pattern repeats. Overall, for the persistent stimulation TNF=1111111111⋯ we obtain NFκB=1010101010⋯. These oscillations agree with the NFκB oscillations reported in [Lipn07] for persistent TNF stimulation.

| $TNF(t)$, $NFκB(t-1)$ | $NFκB(t)$ with normal A20 | $NFκB(t)$ with dysfunctional A20 |
|---|---|---|
| 0, 0 | 0 | 0 |
| 1, 0 | 1 | 1 |
| 0, 1 | 0 | 0 |
| 1, 1 | 0 | 1 |

**Table S1.** Activity states of NFκB at time *t* for normal and abnormal (dysfunctional) A20 conditions in the NFκB network.

In pulsed TNF stimulation, a pulse of TNF is applied to the system only for a short period of time. When this pulse is repeated, as an example, TNF states using our notation can be written as TNF=1000010000⋯. Similarly to the previous paragraph and using table S1, it can be easily shown that NFκB=1000010000⋯. This means that NFκB first becomes activated in response to the TNF pulse, but later becomes inactive due to the lack stimulation, until the next pulse arrives. This behavior agrees with the NFκB data presented in [Lipn07] for pulsed TNF stimulation.

The previous two paragraphs are for the normal NFκB network. With A20 deficiency, the network becomes abnormal. Using our system formulation with dysfunctional A20, in response to TNF=1111111111⋯ we obtain NFκB=1111111111⋯. This indicates long lasting NFκB activity, for persistent TNF stimulation, when A20 is dysfunctional. This behavior agrees with A20-/- experimental cell data in [Lipn07] as well.

**G. Method for calculating communication and computation rates of the NFκB network:** Due to the feedback in the network, the network has a memory such that its output (NFκB) activity state depends on the present and past input (TNF) activity states. Let $Y_1$ and $Y_2$ represent the activity states of NFκB at two consecutive time instants $t=1$ and $t=2$, respectively, i.e., $Y_1 = $ NFκB($t=1$) and $Y_2 = $ NFκB($t=2$). Note that $t=1$ and $t=2$ represent early and late signaling events, respectively. Similarly we have $X_1 = $ TNF($t=1$) and $X_2 = $ TNF($t=2$). Clearly $X$ and $Y$ variables refer to the network input and output in figure 4, respectively. Using the output activity equation NFκB($t$) = TNF($t$) × $\overline{\text{NFκB}(t-1)}$ presented earlier together with NFκB($t=0$) $=0$, the activity state of the *error-free* network output NFκB in terms of the network input TNF at two consecutive time points $t=1$ and $t=2$, i.e., $(Y_1, Y_2) = f(X_1, X_2)$, can be written as listed in table S2.

When A20 is dysfunctional, the network behaves similarly to a linear pathway with no feedback. Therefore, using the output activity equation $\text{NF}\kappa\text{B}(t) = \text{TNF}(t)$ presented earlier when A20 is dysfunctional, the activity state of the *erroneous* network output NFκB in terms of the network input TNF at $t=1$ and $t=2$, i.e., $(Y_1, Y_2) = F(X_1, X_2)$, can be written as listed in table S2. Note that $F \neq f$ due to the last row.

| $(X_1, X_2)$ | Error-free network output sequence $(Y_1, Y_2) = f(X_1, X_2)$, when A20 is normal | Erroneous network output sequence $(Y_1, Y_2) = F(X_1, X_2)$, when A20 is dysfunctional with probability $p$ |
|---|---|---|
| (0,0) | (0,0) | (0,0) |
| (1,0) | (1,0) | (1,0) |
| (0,1) | (0,1) | (0,1) |
| (1,1) | (1,0) | (1,1) |

**Table S2.** Activity states of NFκB at time points $t=1$ and $t=2$, i.e., $(Y_1, Y_2)$, for normal and abnormal (dysfunctional) A20 conditions in the NFκB network, in terms of the activity states of TNF at $t=1$ and $t=2$, i.e., $(X_1, X_2)$.

Consider equi-probable inputs, i.e., $P(X_1=0, X_2=0) = P(X_1=1, X_2=0) = P(X_1=0, X_2=1) = P(X_1=1, X_2=1) = 1/4$. Using the total probability theorem, for the output probabilities we have

$$P(Y_1 = y_1, Y_2 = y_2) = P(Y_1 = y_1, Y_2 = y_2 \mid \text{A20 is normal}) P(\text{A20 is normal}) \\ + P(Y_1 = y_1, Y_2 = y_2 \mid \text{A20 is dysfunctional}) P(\text{A20 is dysfunctional}), \\ = P(Y_1 = y_1, Y_2 = y_2 \mid \text{A20 is normal})(1-p) \\ + P(Y_1 = y_1, Y_2 = y_2 \mid \text{A20 is dysfunctional}) p. \quad (s14)$$

The conditional output probabilities in (s14) can be written as

$$P(Y_1 = y_1, Y_2 = y_2 \mid \text{A20 is normal}) = \sum_{x_1, x_2} P(Y_1 = y_1, Y_2 = y_2 \mid \text{A20 is normal}, X_1 = x_1, X_2 = x_2) \\ \times P(X_1 = x_1, X_2 = x_2), \quad (s15)$$

$$P(Y_1 = y_1, Y_2 = y_2 \mid \text{A20 is dysfunc.}) = \sum_{x_1, x_2} P(Y_1 = y_1, Y_2 = y_2 \mid \text{A20 is dysfunc.}, X_1 = x_1, X_2 = x_2)$$
$$\times P(X_1 = x_1, X_2 = x_2), \quad (s16)$$

where $(x_1, x_2)$ can be $(0,0)$, $(1,0)$, $(0,1)$ or $(1,1)$. Using table S2, the conditional output probabilities in (s15) and (s16) can be calculated for all possible output activity states, i.e., $(y_1, y_2) = (0,0), (0,1), (1,0), (1,1)$, which upon substitution into the second equation of (s14) result in

$$\begin{aligned} P(Y_1 = 0, Y_2 = 0) &= 1/4, \\ P(Y_1 = 0, Y_2 = 1) &= 1/4, \\ P(Y_1 = 1, Y_2 = 0) &= (2-p)/4, \\ P(Y_1 = 1, Y_2 = 1) &= p/4. \end{aligned} \quad (s17)$$

We will use the results in equation (s17) to calculate communication and computation rates of the NFκB network, in the following subsections.

**G1. Communication rate of the NFκB network:** By definition, the communication rate as the mutual information between the network input sequence $(X_1, X_2)$ and the network erroneous output sequence $(Y_1, Y_2)$ is given by

$$R(F) = I(X_1, X_2; Y_1, Y_2) = H(X_1, X_2) - H(X_1, X_2 \mid Y_1, Y_2) = H(Y_1, Y_2) - H(Y_1, Y_2 \mid X_1, X_2), \quad (s18)$$

where $H(.)$ and $H(.\mid.)$ are entropy and conditional entropy, respectively, and $I(.;.)$ is mutual information. They can be calculated using the following expressions

$$H(Y_1, Y_2) = -\sum_{y_1, y_2} P(Y_1 = y_1, Y_2 = y_2) \log_2 P(Y_1 = y_1, Y_2 = y_2), \quad (s19)$$

$$H(Y_1, Y_2 \mid X_1, X_2) = \sum_{x_1, x_2} P(X_1 = x_1, X_2 = x_2) H(Y_1, Y_2 \mid X_1 = x_1, X_2 = x_2), \quad (s20)$$

$$H(Y_1, Y_2 \mid X_1 = x_1, X_2 = x_2) = -\sum_{y_1, y_2} P(Y_1 = y_1, Y_2 = y_2 \mid X_1 = x_1, X_2 = x_2)$$
$$\times \log_2 P(Y_1 = y_1, Y_2 = y_2 \mid X_1 = x_1, X_2 = x_2), \quad (s21)$$

where $\log_2$ is logarithm to the base 2.

By substituting (s17) in (s19) we obtain the following expression for the erroneous network output entropy $H(Y_1, Y_2)$

$$H(Y_1, Y_2) = 2 - 0.25(2-p)\log_2(2-p) - 0.25 p \log_2 p. \quad (s22)$$

On the other hand, using the first three rows of table S2 and equation (s21) it can be verified that $H(Y_1,Y_2 | X_1=0, X_2=0) = H(Y_1,Y_2 | X_1=1, X_2=0) = H(Y_1,Y_2 | X_1=0, X_2=1) = 0$. Moreover, using the last row of table S2 and equation (s21), it can be shown that $H(Y_1,Y_2 | X_1=1, X_2=1) = -(1-p)\log_2(1-p) - p\log_2 p$. By substituting these results in equation (s20) we obtain the following expression for the erroneous network output conditional entropy $H(Y_1,Y_2 | X_1, X_2)$

$$H(Y_1,Y_2 | X_1, X_2) = -0.25((1-p)\log_2(1-p) + p\log_2 p). \quad (s23)$$

Upon substitution of (s22) and (s23) in the second equation of (s18) we finally obtain equation (3) in the paper for the NFκB network communication rate $R(F)$.

Note that we calculate the quantity $I(X_1, X_2; Y_1, Y_2)$ to provide some insights, although it only corresponds to an achievable transmission rate if the dysfunctionality state of A20 changes over time in a memoryless fashion, across pairs of uses of the channel (the NFκB network). This is different from the assumption made that A20 is either normal or dysfunctional for the entire time.

**G2. Computation rate of the NFκB network:** By definition, the computation rate of the network with respect to the error-free output $f$ is given by

$$R_f(F) = H(X_1, X_2) - H(f(X_1, X_2) | Y_1, Y_2), \quad (s24)$$

such that $f(X_1, X_2)$ is the error-free output sequence, while as mentioned previously, $(Y_1, Y_2)$ is the erroneous output sequence $F(X_1, X_2)$. The conditional entropy term $H(f(X_1, X_2) | Y_1, Y_2)$ in (s24) is the amount of information required to specify values of $f(X_1, X_2)$ given the values of $(Y_1, Y_2)$. Since according to table S2, $(Y_1, Y_2)$ values in the third column completely determine $f(X_1, X_2)$ values in the second column, with no ambiguity, the conditional entropy becomes zero, i.e., $H(f(X_1, X_2) | Y_1, Y_2) = 0$. For the network input entropy term $H(X_1, X_2)$ in (s24) and with equiprobable inputs we have

$$H(X_1, X_2) = -\sum_{x_1, x_2} P(X_1=x_1, X_2=x_2)\log_2 P(X_1=x_1, X_2=x_2) = 2. \quad (s25)$$

By substituting these two results in equation (s24), we finally obtain equation (4) in the paper for the NFκB network computation rate $R_f(F)$.

**H. Method for calculating communication capacity of the NFκB network:** To model possible abnormal behavior and erroneous response of the network due to A20 deficiency, we consider A20 as a molecule which has a chance to be dysfunctional with a probability $0 \leq p \leq 1$. More specifically, consider that the probability of A20 to remain 0, inactive, regardless of the signal from NFκB is $p$. This model is consistent with the fact that A20 is inactive in several hematological malignancies [Hymo10]. To calculate the communication capacity $C(F)$, we consider two possible scenarios: normal A20 which corresponds to $p=0$, and dysfunctional A20 which means $0 < p \leq 1$.

**H1. Communication capacity with normal A20:** Let $Y^n$ and $X^n$ represent sequences of length $n$ at the output and input of the NFκB network (figure 4), whose $t$-th elements are $Y_t = \text{NFκB}(t)$ and $X_t = \text{TNF}(t)$, respectively, $t = 1, 2, ..., n$. Due to the A20 feedback in the network, it is a system with memory, i.e., its output NFκB activity state depends on the present and past input TNF activity states. Based on capacity formula for systems with memory [Verd94], the communication capacity of the NFκB network can be written as

$$C(F) = \limsup_{n \to \infty} \sup_{X^n} \frac{1}{n} I(X^n; Y^n) = \limsup_{n \to \infty} \sup_{X^n} \frac{1}{n}(H(X^n) - H(X^n | Y^n)),$$

$$= \limsup_{n \to \infty} \sup_{X^n} \frac{1}{n}(H(Y^n) - H(Y^n | X^n)).$$
(s26)

Here sup stands for supremum, $I(X^n; Y^n)$ is mutual information between the network input and output sequences $X^n$ and $Y^n$, and $H(.)$ and $H(.|.)$ are entropy and conditional entropy, respectively.

To calculate $C(F)$, we obtain upper and lower bounds on the communication capacity. For the upper bound, we note the non-negativity of the conditional entropy, i.e., $H(Y^n | X^n) \geq 0$, which reduces the second equation in (s26) to

$$C(F) \leq \limsup_{n \to \infty} \frac{1}{X^n} \frac{1}{n} H(Y^n). \qquad (s27)$$

To calculate the right-hand side limit in (s27), let $a_n$ be the total number of output sequences $Y^n$ whose lengths are $n$. It is straightforward to note that the total number of output sequences of length $n$ which start with a 0 is $a_{n-1}$. For an output sequence which starts with a 1, we note that the second element has to be 0. This is because we cannot have two consecutive 1s at the network output (see the second column of table S2). Therefore, the total number of output sequences of length $n$ which start with a 1 is $a_{n-2}$. Overall, for the total number of output sequences of length $n$ we have the following relation

$$a_n = a_{n-1} + a_{n-2}, \quad n = 3, 4, \cdots, \qquad (s28)$$

where $a_1 = 2$ and $a_2 = 3$, which correspond to two output sequences 0 and 1, and three output sequences 00, 01 and 10, respectively. Equation (s28) is the Fibonacci sequence [Weis1] for which we have the following closed-form formula

$$a_n = \frac{(\frac{1+\sqrt{5}}{2})^{n+2} - (\frac{1-\sqrt{5}}{2})^{n+2}}{\sqrt{5}}, \quad n = 1, 2, 3, 4, \cdots.$$

(s29)

Since entropy of a variable is less than or equal to the logarithm of the number of elements in its range [Cove91], we can write this inequality for the output entropy $H(Y^n) \leq \log_2 a_n$, which results in

$$\limsup_{n \to \infty} \frac{1}{X^n} \frac{1}{n} H(Y^n) = \lim_{n \to \infty} \frac{1}{n} \log_2 a_n = \log_2 \left( \frac{1+\sqrt{5}}{2} \right). \qquad (s30)$$

By substituting (s30) in (s27), the communication capacity upper bound can be written as

$$C(F) \leq \log_2 \left( \frac{1+\sqrt{5}}{2} \right). \qquad (s31)$$

To obtain the communication capacity lower bound, we consider equi-probable input sequences which do not have two consecutive 1s. In these cases, network outputs will be the same as the network inputs (see the second column of table S2), i.e., $Y^n = X^n$, which means $H(Y^n | X^n) = 0$. By substituting this in the second equation of (s26) and dropping sup we obtain

$$C(F) \geq \lim_{n \to \infty} \frac{1}{n} H(Y^n). \tag{s32}$$

Since the considered input sequences are equi-probable, output sequences are equi-probable as well. Therefore, using the definition of the entropy we obtain

$$H(Y^n) = -\sum_{i=1}^{a_n} P(Y_i^n = y_i^n) \log_2 P(Y_i^n = y_i^n) = -\sum_{i=1}^{a_n} (1/a_n) \log_2 (1/a_n) = \log_2 a_n, \tag{s33}$$

with $a_n$ given in (s29). By substituting (s33) in (s32) and calculating the limit, the communication capacity lower bound can be written as

$$C(F) \geq \log_2 \left( \frac{1+\sqrt{5}}{2} \right). \tag{s34}$$

Overall, the communication capacity of the NFκB network with normal A20 can be obtained by noticing that its upper and lower bounds in (s31) and (s34) are the same, therefore

$$C(F) = \log_2 \left( \frac{1+\sqrt{5}}{2} \right) \approx 0.7, \ p = 0. \tag{s35}$$

Interestingly, the term $(1+\sqrt{5})/2$ is the well-known golden ratio [Weis2], and the Fibonacci sequence appears in a number of natural and biological systems [Wiki]. Another noteworthy observation is that the calculated communication capacity of about 0.7 bits for the TNF-NFκB system is based on early and late responses of NFκB to TNF. It falls between the experimentally-determined individual maximum mutual information of about 0.9 and 0.6 bits for the same system, based on early and late NFκB responses, respectively [Cheo11].

**H2. Communication capacity with dysfunctional A20:** In this case A20 has a likelihood to be dysfunctional with probability $p$, such that $0 < p \leq 1$. Here we consider two scenarios to calculate the communication capacity for the NFκB network: $p = 1$ and $0 < p < 1$.

When A20 is dysfunctional with probability $p=1$, there is no negative feedback in the network as A20 is completely inactive, so, the network in figure 4 becomes a linear pathway. It is straightforward to show the capacity of such a linear pathway is 1 (see "Additional file 1" of [Habi14a] for the method)

$$C(F) = 1, \ p = 1. \tag{s36}$$

When A20 is dysfunctional with the probability of $0 < p < 1$, the NFκB network can be considered to be represented by an erroneous network with probability $p$ and an error-free network with probability $1-p$. Using communication and information theory terminologies, the NFκB network can then be considered as a mixed channel [Han03]. Based on capacity formula for mixed channels [Han03], the communication capacity of the NFκB network can be written as

$$C(F) = \sup_{\mathbf{X}} \min(\tilde{I}(\mathbf{X};\mathbf{Y}_1), \tilde{I}(\mathbf{X};\mathbf{Y}_2)), \tag{s37}$$

in which $\mathbf{X} = \{X^n\}_{n=1}^{\infty}$ is the set of input sequences, whereas $\mathbf{Y}_1 = \{Y_1^n\}_{n=1}^{\infty}$ and $\mathbf{Y}_2 = \{Y_2^n\}_{n=1}^{\infty}$ are the sets of sequences at the outputs of the error-free network and the erroneous network, respectively. Additionally, we have

$$\tilde{I}(\mathbf{X};\mathbf{Y}_i) = \lim_{n\to\infty} \frac{1}{n} I(X^n; Y_i^n) = \lim_{n\to\infty} \frac{1}{n}(H(Y_i^n) - H(Y_i^n \mid X^n)), \ i = 1, 2. \tag{s38}$$

To calculate $C(F)$ in (s37), we obtain upper lower and bounds on the communication capacity. For the upper bound, we note

$$C(F) \leq \min(\max \tilde{I}(\mathbf{X};\mathbf{Y}_1), \max \tilde{I}(\mathbf{X};\mathbf{Y}_2)) = \min\left(\log_2\left(\frac{1+\sqrt{5}}{2}\right), 1\right) = \log_2\left(\frac{1+\sqrt{5}}{2}\right), \tag{s39}$$

where we have used (s35) and (s36), respectively. For the lower bound we can write

$$C(F) \geq \min(\tilde{I}(\mathbf{X};\mathbf{Y}_1), \tilde{I}(\mathbf{X};\mathbf{Y}_2)) = \min\left(\log_2\left(\frac{1+\sqrt{5}}{2}\right), \log_2\left(\frac{1+\sqrt{5}}{2}\right)\right) = \log_2\left(\frac{1+\sqrt{5}}{2}\right), \tag{s40}$$

where the identity is obtained by considering the input probability distribution that achieves the capacity of the first network, i.e., the error-free network. Taken together, the communication capacity of the NFκB network when A20 has a likelihood to be dysfunctional with probability $p$, $0 < p < 1$, can be obtained by noting that its upper and lower bounds in (s39) and (s40) are the same, therefore

$$C(F) = \log_2\left(\frac{1+\sqrt{5}}{2}\right) \approx 0.7, \ 0 < p < 1. \tag{s41}$$

**I. Method for calculating computation capacity of the NFκB network:** For a system with no feedback and no memory, its computation capacity with respect to the error-free output $f$ is given by equation (2) in the paper. For a system with feedback and memory, such as the NFκB network, equation (2) can be extended as follows

$$C_f(F) = \lim_{n\to\infty} \sup_{X^n} \frac{1}{n}(H(X^n) - H(f(X^n)|Y^n)), \tag{s42}$$

in which $f(X^n)$ and $Y^n$ represent error-free and erroneous output sequences of length $n$ of the NFκB network, respectively, whereas $X^n$ is its input sequence of length $n$. To calculate (s42), we consider three possible scenarios for the probability of A20 to be dysfunctional: $p = 0$, $p = 1$ and $0 < p < 1$.

For $p = 0$, normal A20, we have an error-free network where observed outputs are indeed the correct outputs, i.e., $Y^n = f(X^n)$. Based on the definition of the conditional entropy we obtain $H(f(X^n)|Y^n) = H(f(X^n)|f(X^n)) = 0$. On the other hand, since entropy of $X^n$ is less than or equal to the logarithm of the number of elements in its range [Cove91], we can write $H(X^n) \leq \log_2 2^n$, with the equality achieved for equi-probable 0s and 1s. By substituting this and the zero conditional entropy in (s42) we obtain $C_f(F) = 1$ for $p = 0$.

If $p = 1$, completely dysfunctional A20, there is no negative feedback in the network as A20 is completely inactive, so, the network in figure 4 becomes a linear pathway, where observed outputs are actually equal to the inputs, i.e., $Y^n = X^n$. For the conditional entropy this results in $H(f(X^n)|Y^n) = H(f(X^n)|X^n) = 0$, where the last identity is obtained because upon knowing $X^n$, $f(X^n)$ is completely known, without any ambiguity. By substituting this in (s42) and similarly to the previous paragraph, we obtain $C_f(F) = 1$ for $p = 1$.

When $0 < p < 1$, A20 can be dysfunctional with probability $p$ all the time, or can be normal with probability $1-p$ all the time. In the former case, since A20 is inactive and there is no feedback in the network, observed outputs are equal to the inputs, i.e., $Y^n = X^n$, which results in $H(f(X^n)|Y^n) = H(f(X^n)|X^n) = 0$. In the latter case, since the network is error-free, observed outputs are the correct outputs, i.e., $Y^n = f(X^n)$, which results in $H(f(X^n)|Y^n) = H(f(X^n)|f(X^n)) = 0$. Upon substitution of the zero conditional entropy in (s42) and similarly to the previous two paragraphs, we obtain $C_f(F) = 1$ for $0 < p < 1$. Another way of obtaining this result is to note that by looking at the network output sequence, we can always find out the correct outputs. If there are two consecutive 1s at the network output, it means A20 in the network is dysfunctional (see the third column of table S2). As discussed previously, since in this case the observed outputs are equal to the inputs, $Y^n = X^n$, we can find out the inputs, and subsequently recover the correct outputs with no ambiguity, by simply computing $f(X^n)$. On the other hand, if there are no two consecutive 1s at the network output, it means A20 in the network is normal (see the second column of table S2). In this case the observed outputs are directly giving us the correct outputs $f(X^n)$, as the network is error-free now.

Overall, the computation capacity of the NFκB network with respect to the error-free output $f$ is

$$C_f(F) = 1, \ 0 \leq p \leq 1. \tag{s43}$$

**J. Capacity definitions and formulas:** Consider a system such as a signaling network, with $X$ as its input, which computes the output according to the error-free mapping $f$. So, the error-free output is $f(X)$. When the system is erroneous due to the presence of some dysfunctional molecules, the mapping is called $F$, so, the erroneous output is $F(X)$. If we consider the system as a communication channel, its communication capacity can be informally expressed as

$$C(F) = \log_2(\text{max number of inputs } X \text{ such that } X \text{ can be correctly recovered from } F(X)),$$
$$= \log_2(\text{max number of decodable inputs}). \quad \text{(s44)}$$

Formally speaking, the communication capacity formula is

$$C(F) = \max_{P(X)} \{H(X) - H(X|F(X))\}. \quad \text{(s45)}$$

In contrast, if we consider the system as a computing machine, its computation capacity can be informally defined as

$$C_f(F) = \log_2(\text{max number of inputs } X \text{ such that } f(X) \text{ can be correctly recovered from } F(X)),$$
$$= \log_2(\text{max number of computable inputs}). \quad \text{(s46)}$$

Equation (s46) can be re-written as

$$C_f(F) = \log_2(\text{max no. of } f(X) \text{ values that can be correctly recovered from } F(X) \times |f^{-1}(Z)|),$$
$$= \log_2(\text{max no. of } f(X) \text{ values that can be correctly recovered from } F(X))$$
$$+ \log_2(|f^{-1}(Z)|), \quad \text{(s47)}$$
$$= \text{communication capacity of virtual channel } \Phi \text{ between } f \text{ and } F + \log_2(|f^{-1}(Z)|),$$

where $Z = f(X)$ and $|f^{-1}(Z)|$ is the number of elements in the inverse image of $Z$ through the error-free function $f$. The virtual channel or transformation $\Phi$ defined between $f$ and $F$, and the $f^{-1}$ notation are illustrated in the transition probability diagram, figure S2(c), of the malfunctioning system numerical example of the next section.

We will prove in Section L2 of Supplementary Material that the computation capacity is given as

$$C_f(F) = \max_{P(X)} \{I(f(X); F(X)) + H(X|f(X))\}. \quad \text{(s48)}$$

In equation (s48), $I(f(X); F(X))$ is the mutual information between $f(X)$ and $F(X)$, whose maximization results in the communication capacity of the virtual channel $\Phi$ defined between $f$ and $F$, whereas the second term of (s48) is obtained by noting that entropy of a random variable is less than or equal to the logarithm of the number of elements in its range [Cove91], i.e., $H(X|f(X)) \leq \log_2(|f^{-1}(Z)|)$. Upon replacing $I(f(X); F(X))$ in (s48) by its definition we obtain

$$C_f(F) = \max_{P(X)} \{H(f(X)) - H(f(X)|F(X)) + H(X|f(X))\}. \quad \text{(s49)}$$

Based on a property of entropy and conditional entropy [Cove91], we have $H(X) - H(X | f(X)) = H(f(X)) - H(f(X) | X)$. Note that $H(f(X) | X) = 0$ since $f$ is a deterministic function. This indicates $H(f(X)) + H(X | f(X)) = H(X)$, that upon substitution in equation (s49), results in the computation capacity formula

$$C_f(F) = \max_{P(X)} \{H(X) - H(f(X)|F(X))\}. \quad \quad (s50)$$

Based on the data processing inequality, we have $H(f(X)| F(X)) < H(X| F(X))$ (the two conditional entropies are equal in the special case of $f$ being an invertible function, i.e., if and only if $f$ is a one-to-one function). Upon comparing equation (s50) with (s45), we conclude that $C_f(F) > C(F)$. Intuitively, this means that it is easier to correctly compute a non-invertible (not one-to-one) function than to decode it correctly.

**K. An example of a malfunctioning system and its communication and computation capacities:** Consider a system (figure S2a) that in response to eight different input values $\{x_1,...,x_8\}$, generates four different output values $\{\alpha, \beta, \gamma, \delta\}$, according to the following error-free mapping $f$ (with no dysfunctional component in the system)

$$\begin{aligned} f(x_1) &= f(x_2) = \alpha, \\ f(x_3) &= f(x_4) = \beta, \\ f(x_5) &= f(x_6) = \gamma, \\ f(x_7) &= f(x_8) = \delta. \end{aligned} \quad \quad (s51)$$

Note that $x_1,...,x_8$ and $\alpha, \beta, \gamma, \delta$ are all real-valued numbers. The error-free system response is graphically shown in the left half of figure S2c.

When some system components become dysfunctional, the system turns into a malfunctioning one that computes some outputs incorrectly (figure S2b). Here we consider the following numerical example

$$F(x_1) = \alpha \text{ or } \beta \text{ with probability } 1/2, \quad F(x_2) = \alpha \text{ or } \beta \text{ with probability } 1/2,$$
$$F(x_3) = \alpha \text{ or } \beta \text{ with probability } 1/2, \quad F(x_4) = \alpha \text{ or } \beta \text{ with probability } 1/2,$$
$$F(x_5) = \gamma \text{ or } \delta \text{ with probability } 1/2, \quad F(x_6) = \gamma \text{ or } \delta \text{ with probability } 1/2, \quad \text{(s52)}$$
$$F(x_7) = \gamma \text{ or } \delta \text{ with probability } 1/2, \quad F(x_8) = \gamma \text{ or } \delta \text{ with probability } 1/2.$$

This erroneous system response is graphically depicted in the right half of figure S2c. The solid and dashed arrows represent correct and incorrect transitions, respectively, with their probabilities provided next to the arrows. This diagram reflects that, for example, $F(x_1) = \alpha$ is a correct response, whereas $F(x_1) = \beta$ is an incorrect response. The transition probabilities (figure S2c) are basically conditional probabilities, as defined below

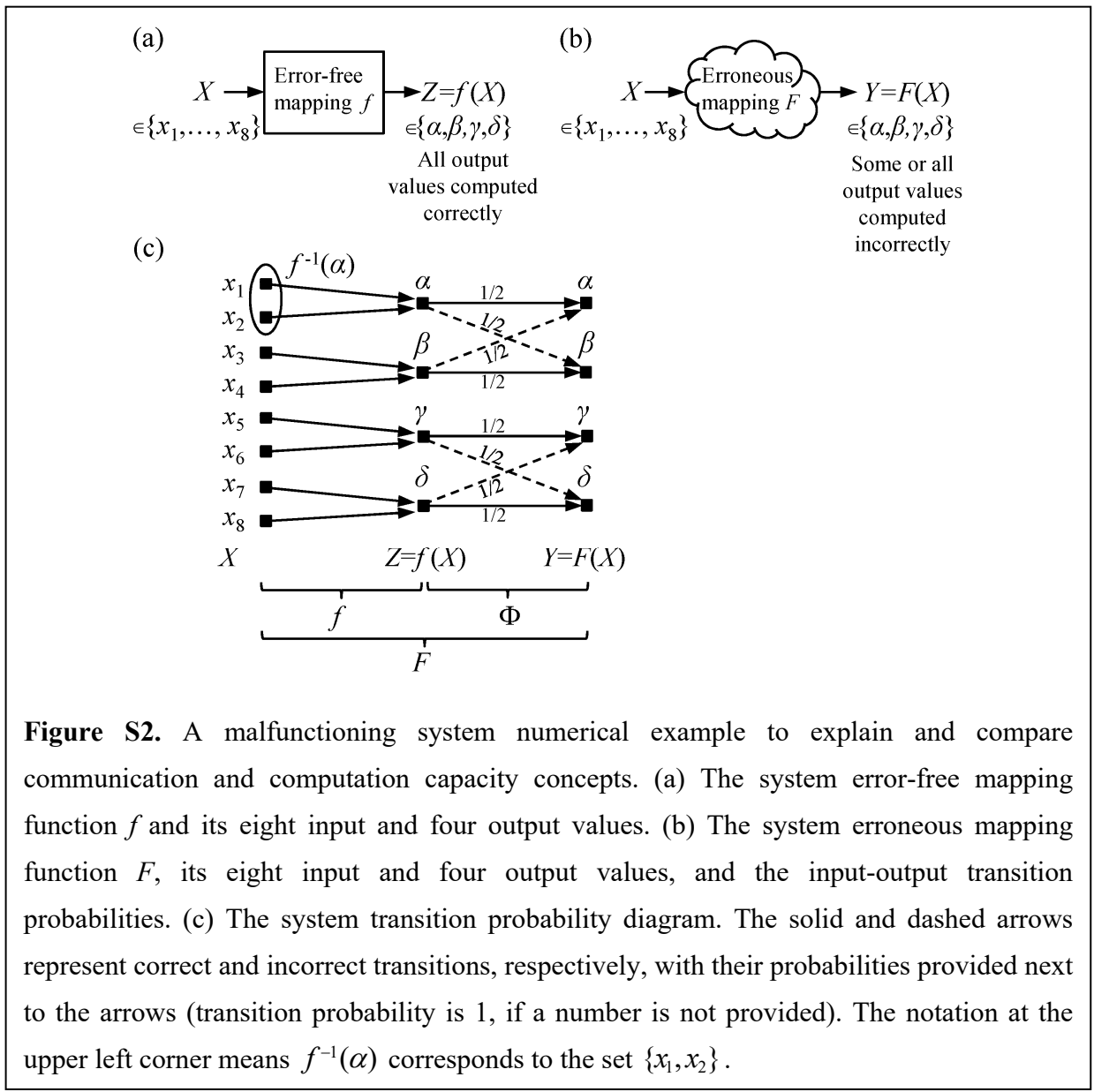

**Figure S2.** A malfunctioning system numerical example to explain and compare communication and computation capacity concepts. (a) The system error-free mapping function $f$ and its eight input and four output values. (b) The system erroneous mapping function $F$, its eight input and four output values, and the input-output transition probabilities. (c) The system transition probability diagram. The solid and dashed arrows represent correct and incorrect transitions, respectively, with their probabilities provided next to the arrows (transition probability is 1, if a number is not provided). The notation at the upper left corner means $f^{-1}(\alpha)$ corresponds to the set $\{x_1, x_2\}$.

$$P(Y = y \mid Z = z) = \begin{cases} 1/2, & y \in \{\alpha, \beta\}, \text{ if } z = \alpha \text{ or } \beta, \\ 1/2, & y \in \{\gamma, \delta\}, \text{ if } z = \gamma \text{ or } \delta. \end{cases} \quad (s53)$$

*Communication capacity $C(F)$ of the malfunctioning system (figure S2b) with the given transition probability diagram (figure S2c):* Transition probability channel matrix $\mathbf{M}_{Y|X}$ of a system with input $X$ and output $Y$ is a matrix whose $x$-th row and $y$-th column is the conditional probability $P(Y = y \mid X = x)$. According to the system transition probability diagram (figure S2c), the transition probability channel matrix for the malfunctioning system (figure S2b) can be written as

$$\mathbf{M}_{Y|X} = \begin{bmatrix} 1/2 & 1/2 & 0 & 0 \\ 1/2 & 1/2 & 0 & 0 \\ 1/2 & 1/2 & 0 & 0 \\ 1/2 & 1/2 & 0 & 0 \\ 0 & 0 & 1/2 & 1/2 \\ 0 & 0 & 1/2 & 1/2 \\ 0 & 0 & 1/2 & 1/2 \\ 0 & 0 & 1/2 & 1/2 \end{bmatrix}. \quad (s54)$$

Since all rows of $\mathbf{M}_{Y|X}$ are permutations of each other, and all columns of $\mathbf{M}_{Y|X}$ are permutations of each other, $\mathbf{M}_{Y|X}$ represents a symmetric channel [Cove91]. Communication capacity of a symmetric channel is: $\log_2(\text{number of output values}) - \text{entropy of one row of } \mathbf{M}_{Y|X}$ [Cove91]. For these two terms in our example we have $\log_2(4) = 2$ and $-(1/2)\log_2(1/2) - (1/2)\log_2(1/2) - 0\log_2(0) - 0\log_2(0) = 1$, respectively (note that by definition, $0\log_2(0) = 0$). Therefore, the system communication capacity is $C(F) = 1$ bit.

*Computation capacity $C_f(F)$ of the malfunctioning system (figure S2b) with the given transition probability diagram (figure S2c):* First we informally calculate $C_f(F)$ using the last expression in equation (s47). For its first term, i.e., "communication capacity of the virtual channel Φ between $f$ and $F$" (figure S2c), we note that its transition probability channel matrix $\mathbf{M}_{Y|Z}$ with input $Z$ and output $Y$, using the transition probability diagram (figure S2c), can be written as

$$\mathbf{M}_{Y|Z} = \begin{bmatrix} 1/2 & 1/2 & 0 & 0 \\ 1/2 & 1/2 & 0 & 0 \\ 0 & 0 & 1/2 & 1/2 \\ 0 & 0 & 1/2 & 1/2 \end{bmatrix}. \qquad (s55)$$

Since $\mathbf{M}_{Y|Z}$ represents a symmetric channel, using the previously explained communication capacity formula for such channels [Cove91], the communication capacity of the virtual channel $\Phi$ between $f$ and $F$ can be verified to be: $\log_2(\text{number of output values}) - \text{entropy of one row of } \mathbf{M}_{Y|Z} = 2 - 1 = 1$ bit. For the second term of the last expression in equation (s47), we note that it is equal to $\log_2(2) = 1$ bit, because $|f^{-1}(Z)| = 2$ for each $Z$ from the set $\{\alpha, \beta, \gamma, \delta\}$ (see, for example, upper left corner of figure S2c). By substituting these results in the last formula of equation (s47), the system computation capacity is obtained to be $C_f(F) = 1 + 1 = 2$ bits.

To formally calculate $C_f(F)$, using equation (s48), we note that by definition, $I(f(X); F(X)) \leq $ communication capacity of the virtual channel between $f$ and $F$, where the latter was shown after equation (s55) to be 1 bit, therefore, $I(f(X); F(X)) \leq 1$. Additionally, as stated at the end of the paragraph immediately after equation (s48), $H(X|f(X)) \leq \log_2(|f^{-1}(Z)|) = \log_2(2) = 1$, $\forall Z \in \{\alpha, \beta, \gamma, \delta\}$. Overall, we have $I(f(X); F(X)) + H(X|f(X)) \leq 1 + 1 = 2$, where the equality can be achieved using a uniform distribution over four input values, for example, $\{x_1, x_2, x_5, x_6\}$ (by selecting these four values, the correct output $f(x)$ can be unambiguously obtained from $F(x)$). Therefore, according to equation (s48) we finally obtain $C_f(F) = 2$ bits. Note that it is greater than $C(F) = 1$ bit, as expected, because $f$ is not a one-to-one function (as graphically shown in the left half of figure S2c).

Intuitively speaking, the maximum number of correctly computable inputs in this example is $2^{C_f(F)} = 4$, whereas the maximum number of correctly decodable or distinguishable inputs is $2^{C(F)} = 2$.

**L. Communication and computation coding theorems for malfunctioning systems:** In this section, first we review a fundamental coding theorem for error-causing communication channels [Cove91], that highlights the implications of the communication capacity concept. Then we present and prove a similar coding theorem for error-causing computing machines [Simo10], to elucidate the computation capacity concept and its significance. To simplify the notation, we use $C$ and $C_f$ for communication and computation capacities, respectively, instead of $C(F)$ and $C_f(F)$.

**L1. Communication capacity:** To reliably transmit (communicate) $M$ values $\{x_1,...,x_M\}$ of the input $X$ through an error-causing channel, as taught by Shannon [Cove91], one needs to design and transmit $M$ codewords instead, each of length $n$. The codeword $x_m^n$ is assigned by an encoder to the $m$-th value $x_m$, and contains $n$ symbols, i.e., $x_m^n = (x_{m,1},...,x_{m,n})$, with $n$ being as large as needed. To transmit $x_m$, $n$ values of the codewrod $x_m^n$ are transmitted one after the other ($n$ transmissions). The communication rate of this ($M,n$) code is $R = \log_2(M)/n$ bits per channel use.

In a memoryless channel, for each transmitted value $x_{m,i}$, $i=1,...,n$, the receiver obtains a possibly different version of $x_{m,i}$, called $y_i = F(x_{m,i})$. The receiver collects all the $y_i$ values into a vector $y^n = (y_1,...,y_n)$, and carries out decoding with the goal of identifying the transmitted codeword $x_m^n$, or, equivalently, the originally transmitted $x_m$ value of the input $X$. The decoding function $g(y^n)$, which returns a value in the set $\{x_1,...,x_M\}$, can be interpreted as partitioning the space of received values into $M$ decision regions. If $x_m$ is transmitted through the channel and after decoding at the receiver, it turns out that $g(y^n) = x_m$, transmission is successful and error free, otherwise, it is erroneous.

Here is a simple example to better understand the process. To transmit $x_1$, the encoder generates and transmits the codeword $x_1^n$ of length $n$ through the communication channel. The received codeword of length $n$ is $y^n$, which upon decoding results in $g(y^n)$. If $g(y^n) = x_1$, the transmission is successful and error free, i.e., correct recovery or decoding or distinguishing $x_1$, whereas if $g(y^n) = x_2$ or ... or $x_M$, the transmission is unsuccessful and erroneous. The associated conditional decoding error probabilities are $\lambda_m = P(g(Y^n) \neq x_m \mid X = x_m)$, $m = 1,..., M$, and the maximal decoding error probability for an (M,n) code is defined as $\lambda^{(n)} = \max \lambda_m$, $m = 1, ..., M$. With these definitions, the fundamental coding theorem for discrete memoryless communication channels can be written as follows [Cove91]

*Communication channel coding theorem:* In a discrete memoryless communication channel, for any rate $R < C$, reliable communication with arbitrarily low error probability is possible, i.e., there exists a sequence of $(2^{nR}, n)$ codes such that the maximal decoding error probability $\lambda^{(n)} \to 0$ as $n \to \infty$ (rate achievability). Conversely, any sequence of $(2^{nR}, n)$ codes with $\lambda^{(n)} \to 0$, as $n \to \infty$, must satisfy $R < C$ (converse property). This means for any rate $R > C$, reliable communication with arbitrarily low error probability is not possible, i.e., $\lambda^{(n)}$ does not approach 0.

**L2. Computation capacity:** Now we turn our attention to a coding theorem for error-causing computing machines [Simo10], which clarifies the computation capacity concept and what it implies. The reliable computation goal is to compute a function $z = f(x)$ reliably from its erroneous version $y = F(x)$, for a set of input values $x$ selected uniformly at random within a given set. The computation capacity aims at quantifying the number $M$ of distinct inputs for which an observer of the possibly incorrect output value $y = F(x)$, can recover the desired function value $z = f(x)$. In other words, the computation capacity quantifies the number of *computable* inputs, rather than the number of inputs that can be correctly communicated.

To elaborate on this notion, it is noted that a constant function has a large computation capacity, since the observer can always recover the correct (constant) output from an erroneous output. In contrast, a function with significant variations has generally a smaller capacity in the presence of errors. These considerations suggest that the size of pre-images or inverse images $f^{-1}(z)$ of the function *f* plays an important role in the definition of computation capacity: the more values of $x$ are mapped to the same value $z$ of the function $z = f(x)$, i.e., the larger the sets $f^{-1}(\cdot)$ are, the more input values can be computed on, by recovering the corresponding value of the function.

More precisely, to compute $M$ function values $\{f(x_1), ..., f(x_M)\}$ reliably from their erroneous versions $\{F(x_1), ..., F(x_M)\}$, for a set of input values $\{x_1, ..., x_M\}$, one needs an encoder. The encoder can generate $M$ codewords, each of length $n$, i.e., $x_m^n = (x_{m,1}, ..., x_{m,n})$, $m = 1, ..., M$, with $n$ being as large as needed and each codeword is selected uniformly at random. To compute the function value at $x_m$, $n$ function values of the codeword $x_m^n$ are computed one after the other ($n$ computations). The computation rate of this ($M$,$n$) code is $R_f = \log_2(M)/n$ bits per function use.

We assume a memoryless model whereby the erroneous function values $y_i = F(x_{m,i})$, $i = 1, ..., n$, are obtained from the corresponding correct values $z_i = f(x_{m,i})$, $i = 1, ..., n$, through the independent application of an error-causing channel (see, for example, $\Phi$ in figure S2c). Therefore, the computed erroneous values $\{y_i = F(x_{m,i})\}$ are conditionally independent, given the correct function values $z^n = f^n(x_m^n) = (f(x_{m,1}), ..., f(x_{m,n}))$. The observer collects all erroneous outputs into a vector $y^n = F^n(x_m^n) = (F(x_{m,1}), ..., F(x_{m,n}))$ and carries out decoding with the goal of identifying the correct function outputs $f^n(x_m^n) = (f(x_{m,1}), ..., f(x_{m,n}))$. Importantly, this task is different from identifying the sequence of inputs, or equivalently the codeword $x_m^n$, which is the goal in the communication framework.

The function $f(x)$ is not generally one-to-one (or bijective), but rather multiple values of the input $x$ are mapped to the same value $f(x)$ (surjective function). For any sequence of correct values $z^n$ of the function, define the set of all inputs $x_m^n$ that map to it as $f^{-n}(z^n) = \{x_m^n: f(x_{m,i}) = z_i\}$. Note that this is the pre-image of the desired function $f$ when applied $n$ times, and that the set of all these pre-images partition the input space. Each input sequence $x_m^n$ maps to a sequence $f^n(x_m^n)$, and the set of all such distinct values is equal to the number of pre-images $f^{-n}(z^n)$ that intersect with the set of input sequences. We index the set of distinct values of $z_k^n$, as the input codeword varies in the codebook $\{x_m^n, m=1,...,M\}$, according to $k = 1, ..., K \leq M$, where equality holds if and only if the function $f$ is one-to-one.

When the correct function value sequence is $z_k^n$, if the decoding function gives $g(y^n) = k$, then computation is successful and error free, while it is otherwise erroneous. The associated conditional decoding error probabilities are $\lambda_k = P(g(Y^n) \neq k \mid \text{input is in set } f^{-n}(z_k^n))$, $k = 1, \ldots, K < M$ (none one-to-one $f$), and the maximal decoding error probability for an $(M,n)$ code is defined as $\lambda^{(n)} = \max \lambda_k$, $k = 1, \ldots, K < M$. With these definitions, we have the following coding theorem for discrete-alphabet memoryless functions and computing machines [Simo10]

*Computing machine coding theorem:* For a discrete memoryless computing machine, for any rate $R_f < C_f$, reliable computation of $f$ with arbitrarily low error probability is possible, i.e., there exists a sequence of $(2^{nR_f}, n)$ codes such that the maximal decoding error probability $\lambda^{(n)} \to 0$ as $n \to \infty$ (rate achievability). Conversely, any sequence of $(2^{nR_f}, n)$ codes with $\lambda^{(n)} \to 0$, as $n \to \infty$, must satisfy $R_f < C_f$ (converse property). This means for any rate $R_f > C_f$, reliable computation with arbitrarily low error probability is not possible, i.e., $\lambda^{(n)}$ does not approach 0.

*Proof:* The rate achievability can be proved using random coding [Cove91] and the error probability analyzed for the virtual communication channel between $f$ and $F$ (see [Simo10] for details).

To prove the converse property, we note that there are $M = 2^{nR_f}$ codewords $X^n$, which if are uniformly selected at random, result in $H(X^n) = nR_f$, using the definition of entropy. By adding $\mp\{H(f^n(X^n)) + H(f^n(X^n)|F^n(X^n))\}$ to $H(X^n)$ we obtain

$$nR_f = \underbrace{H(X^n) - H(f^n(X^n))}_{(a)} \underbrace{- H(f^n(X^n)|F^n(X^n)) + H(f^n(X^n))}_{(b)} + \underbrace{H(f^n(X^n)|F^n(X^n))}_{(c)}. \quad (s56)$$

According to the paragraph under equation (s49), expression (a) is equal to $H(X^n \mid f^n(X^n))$, whereas expression (b) is nothing but $I(f^n(X^n); F^n(X^n))$, i.e., the mutual information between $f^n(X^n)$ and $F^n(X^n)$, and expression (c) will be discussed later. Upon substitution of these identities, $nR_f$ in equation (s56) reduces to

$$nR_f = \underbrace{H(X^n \mid f^n(X^n))}_{(a)} + \underbrace{I(f^n(X^n); F^n(X^n))}_{(b)} + \underbrace{H(f^n(X^n)|F^n(X^n))}_{(c)}. \quad (s57)$$

Since entropy of a set of random variables is less than or equal to the sum of individual entropies [Cove91], we obtain $H(X^n | f^n(X^n)) \leq \sum_{i=1}^{n} H((X^n)_i | f((X^n)_i))$ for expression (a) in (s57). Additionally, since for a discrete memoryless channel, *n*-term joint mutual information between input and output vectors is less than or equal to the sum of *n* individual mutual information terms [Cove91], for the virtual channel between $f$ and $F$, we can write $I(f^n(X^n); F^n(X^n)) \leq \sum_{i=1}^{n} I(f((X^n)_i); F((X^n)_i))$ for expression (b) in (s57). Moreover, based on vanishing error probabilities in the converse property assumption and using Fano's inequality [Cove91], we have $H(f^n(X^n)|F^n(X^n)) \leq n\varepsilon_n$ for expression (c) in (s57), such that $\varepsilon_n \to 0$ as $n \to \infty$. Upon substituting these inequalities in equation (s57), we obtain the following upper bound for $nR_f$

$$nR_f \leq \sum_{i=1}^{n} \left\{ H((X^n)_i | f((X^n)_i)) + I(f((X^n)_i); F((X^n)_i)) \right\} + n\varepsilon_n. \qquad (s58)$$

Based on the definition of the computation capacity $C_f$ in equation (s48), the term within the curly brackets in (s58) is upper bounded by $C_f$. Therefore

$$nR_f \leq nC_f + n\varepsilon_n. \qquad (s59)$$

Since $\varepsilon_n \to 0$ as $n \to \infty$, the inequality in (s59) reduces to $R_f \leq C_f$. This completes the proof of the converse property.

The above theorem can be extended to systems with memory using standard information-theoretic tools (see, e.g., [Cove91]). It can also be generalized to mixed channels, such as the one considered in Section H2 of Supplementary Material.